\documentclass[twocolumn, trackchanges]{aastex63}

\usepackage{graphicx}  
\usepackage{textcomp}
\usepackage{longtable}
\usepackage{threeparttable}
\usepackage{amsmath}
\turnoffedit

\begin{document}

\title{\edit1{Variability, periodicity and contact binaries in WISE}}

\author[0000-0003-2427-4287]{Evan Petrosky}
\affiliation{Department of Physics \& Astronomy, Johns Hopkins University, 3400 N Charles St, Baltimore, MD 21218, USA}

\author[0000-0003-4250-4437]{Hsiang-Chih Hwang}
\affiliation{Department of Physics \& Astronomy, Johns Hopkins University, 3400 N Charles St, Baltimore, MD 21218, USA}

\author[0000-0001-6100-6869]{Nadia L. Zakamska}
\affiliation{Department of Physics \& Astronomy, Johns Hopkins University, 3400 N Charles St, Baltimore, MD 21218, USA}

\author[0000-0002-0572-8012]{Vedant Chandra}
\affiliation{Department of Physics \& Astronomy, Johns Hopkins University, 3400 N Charles St, Baltimore, MD 21218, USA}

\author[0000-0001-7493-2167]{Matthew J. Hill}
\affiliation{Knovita, 1405 Point Street, Baltimore, MD 21231, USA}


\begin{abstract}

The time-series component of WISE is a valuable resource for the study of \edit1{variable objects}.  We present an analysis of an all-sky sample of $\sim$450,000 AllWISE+NEOWISE infrared light curves of likely variables identified in AllWISE.  By computing periodograms of all these sources, we identify $\sim$56,000 periodic variables. Of these, $\sim$42,000 are short-period ($P<1$ day), near-contact or contact eclipsing binaries, many of which are on the main sequence.  We use the periodic and aperiodic variables to test computationally inexpensive methods of periodic variable classification and identification, utilizing various measures of the probability distribution function of fluxes and of timescales of variability.  \edit1{The combination} of variability measures from our periodogram and non-parametric analyses with infrared colors \edit1{from WISE} and absolute magnitudes, colors and \edit1{variability amplitude} from Gaia \edit1{is useful} for the identification and classification of periodic variables.  Furthermore, we show that the effectiveness of non-parametric methods for the identification of periodic variables is comparable to that of the periodogram but at a much lower computational cost.  Future surveys can utilize these methods to accelerate more traditional time-series analyses and to identify evolving sources missed by periodogram-based selections. 
\end{abstract}  

\keywords{binaries: close -- binaries: eclipsing -- binaries: general -- catalogues -- methods: statistical -- stars: variables: general}

\section{Introduction} \label{sec:intro}

The study of the variable sky can yield a wealth of information on a wide range astronomical objects such as asteroids, exoplanets, stars, and active galactic nuclei.  As a result of this broad importance, over the last 20 years, there has been significant investment in high-cadence variability surveys both on the ground (OGLE -- \citealt{udal03}; ASAS -- \citealt{ruci06}; Catalina -- \citealt{drak09}; ASAS-SN -- \citealt{shap14, koch17, jaya18}; ZTF -- \citealt{bell19}; LSST -- \citealt{ivez19}) and in space (Kepler -- \citealt{boru10, koch10}; TESS -- \citealt{rick15}; Gaia -- \citealt{prus16, gaia18a}). 

Considerable effort goes into the sorting of data from these surveys to first differentiate between variable and non-variable objects and then to classify different types of variability.  Traditionally, variable objects are classified based on the similarity of their light curves and colors to known variable prototypes \citep{eyer19a}.  Often, a time-series analysis tool such as a periodogram is run on objects displaying variability to differentiate between periodic and aperiodic variables.  This step requires a clear understanding of the effects of the observing cadence on period recovery and a careful weighing of the pros and cons of different period-search algorithms.  The results of the periodogram-based analysis are often taken, together with measures of light curve morphology and other characteristics of the source (e.g. color) and used as inputs into a classifier (e.g. \citealt{debo07, debo09, sarr09, rich11, duba11, rich12, masc14, kim16, jaya19a, jaya19b, eyer19b}).  

Periodograms, light curve fitting, and machine learning classification are potent tools and will continue to be important to the astronomical community.  Nevertheless, one downside of these methods is that they are often survey-dependent, difficult to physically interpret, and involve a lengthy learning curve and a lot of computational power in order to implement.  To help mitigate this problem, some previous works have used non-parametric variability measures (e.g. \citealt{kine06, pala13, drak13, rimo13, drak14a, drak17, torr15, hill15, find15}).  

In this work we are particularly interested in eclipsing stellar binaries.  Binary stars undergo interesting evolution \citep{step95, fabr07, ivan13, duch13, bork16, moe17, hwan20} and are relevant for a wide array of astronomical phenomena.  They have been used in cosmology as distance indicators \citep{ries11}, in stellar astrophysics as testing grounds for precision stellar evolutionary models \citep{piet04}, and can even host planets \citep{doyl11}.  Binary systems have been linked to transients such as Luminous Red Novae \citep{tyle11, kasl12} and are thought to serve as progenitors for some of the most fascinating objects in the universe -- ultra-compact binaries of white dwarfs, neutron stars and black holes, the tantalizing source population for type Ia supernovae, kilonovae, and gravitational waves \citep{weis10, knig11, maoz14, post14, brow16, belc16, smar17, cowp17, abbo17, temm20}.  Many surveys have associated eclipsing binary (EB) catalogs (e.g.  ASAS -- \citealt{pacz06}; Kepler -- \citealt{kirk16}; Catalina -- \citealt{drak14a}; OGLE -- \citealt{sosz16}; ASAS-SN -- \citealt{jaya19a}).  Future Gaia data releases will include an all-sky EB catalog and the current release identifies a variety of other types of variability \citep{eyer18, holl18, roel18, moln18, mowl18, eyer19a, rimo19, clem19, siop20}.

In this work, we use non-parametric light curve analysis techniques in conjunction with more traditional time-series analysis methods to study variability and identify a sample of eclipsing binaries in data from the Wide-Field Infrared Survey Explorer (WISE; \citealt{wrig10, main11}). The all-sky and long-term coverage, decreased effects of extinction compared to optical wavelengths, and non-uniform cadence probing a wide range of variability timescales \citep{hoff12} make WISE a unique probe of Galactic variability.  Previously, \citet{chen18} used WISE to identify $\sim$50,000 periodic variable candidates of which $\sim$ 42,000 were binaries. Here, we expand on this work and present an analysis of $\sim$450,000 light curves of variables selected using AllWISE variability metrics based on r.m.s. flux variations.  Our larger period search grid allows us to detect short-period objects that were missed by \citet{chen18}.  We also employ a different period-search algorithm, use data from a more recent release, and cross-match our results with Gaia DR2.  In the end, we identify an all-sky sample of $\sim$ 56,000 periodic variable candidates, of which $\sim$ 51,000 are binaries.  We also present the calculation of various non-parametric variability measures for the remaining 394,000 aperiodic variables. 

In Section \ref{sec:WISE_data}, we discuss the WISE data and our time-series analysis. In Section \ref{sec:cont} we explore the contents of our periodic variable selection and display the results of various cross-matches.  In Section \ref{sec:nonparm} we introduce our non-parametric measures and use them to classify periodic variables. In Section \ref{sec:disc} we discuss main-sequence (MS) binaries, short-period objects, extra-galactic and young-stellar-object variability, and the application of non-parametric methods to the identification of periodic variables.  We conclude in Section \ref{sec:conc}. Throughout this paper, WISE magnitudes are quoted on the Vega system, and the following conversions apply: ${\rm W}i_{\rm AB}={\rm W}i_{\rm Vega}+\Delta m_i$, with $\Delta m_i=(2.699, 3.339, 5.174, 6.620)$\footnote{\url{https://wise2.ipac.caltech.edu/docs/release/allsky/expsup/sec4_4h.html}}.

\section{WISE Data and Method}\label{sec:WISE_data}

\subsection{WISE Mission}

The Wide-field Infrared Survey Explorer (WISE) was launched in December 2009 and conducted observations of the entire sky in bands centered on 3.4, 4.6, 12 and 22 $\micron$ (W1, W2, W3, and W4 respectively) until \edit1{the end of September 2010} when it ran out of coolant \citep{wrig10, main11}. From \edit1{October 2010 to February 2011} it conducted observations in the 3.4 and 4.6 $\micron$ bands as a part of the NEOWISE post-cryogenic mission \citep{main11}.  The spacecraft was in hibernation from the end of the post-cryogenic mission until \edit1{December} 2013 when it was reawakened as a part of the NEOWISE Reactivation (NEOWISE-R) mission \citep{main14}. The AllWISE data release includes data from both the original mission and the post-cryogenic mission as of 2013\footnote{\url{http://wise2.ipac.caltech.edu/docs/release/allwise/}}, whereas the NEOWISE Reactivation 2019 Data Release includes all of the data from the time that the spacecraft was awakened from hibernation until 2019.

The WISE spacecraft orbits the Earth with a period of $\sim$5700 seconds ($\sim$0.066 days=1.6 hours)\footnote{\url{http://wise2.ipac.caltech.edu/docs/release/allsky/expsup/sec1_1.html}} on a polar orbit -- near the dividing line between night and day -- and always looks away from the Earth.  Every six months, it images the same portion of the sky and obtains at least eight passes on each point of sky due to the partial overlap of the field of view on consecutive orbits. The sources outside the ecliptic are more frequently observed. The cadence of the data is such that every six months there is a collection of $\sim$10 data points that are each spaced one orbit apart. 

The data pre-processing and extraction of the light curves are different for the AllWISE and NEOWISE releases. AllWISE stacks all the scans, identifies the objects, and then measures the per-scan magnitude of each object at a fixed position. In contrast, NEOWISE identifies objects and measures their photometry in individual scans without stacking. \citet{main14} find  systematic changes in W1 between AllWISE and NEOWISE-R to be $0.01$ magnitude for sources with 8$<$W1$<$14 mag and on the order of $0.1$ magnitude for sources with 14$<$W1$<$15 mag. Outside this range there are magnitude dependent systematic offsets between AllWISE and NEOWISE data.  We limit our \edit1{study} to 8$<$W1$<$15 mag to ensure concordant AllWISE and NEOWISE measurements and to exclude saturated sources \citep{niku14, main14}. We also explicitly apply a cut to ensure that the difference between the mean magnitude of AllWISE and NEOWISE be less than $0.15$ mag because the time-series analysis is not reliable in the case of a large magnitude offset.  A typical individual exposure for sources in the range 8$<$W1$<$15 mag has an uncertainty of $\sim$ $0.03$ mag.  \edit1{A plot of the r.m.s. of single exposure flux measurements versus W1 magnitude is shown in Figure 8 of \citet{main14}.  In the range 8$<$W1$<$15 mag, individual data point uncertainties tend to increase as a function of W1 magnitude.} We incorporate these error measurements into our analysis.

\subsection{WISE Light Curves}

WISE reports a measure of the flux variation in each band based on AllWISE single-epoch photometry.  Each source is assigned a single-digit \texttt{var\_flg} ranging from 0 to 9 in each band, such that the probability that the object's flux does not vary in said band is $\propto 10^{-{\rm var\_flag}}$ \citep{hoff12}. 

We select WISE variable sources having W1 \texttt{var\_flg} $\geq$ 6.  \edit1{We find that that proportion of variable objects found to be periodic via periodogram analysis drops steadily as \texttt{var\_flg} decreases from $\sim$23\% for sources with \texttt{var\_flg} == 9 to $\sim$5\% for sources with \texttt{var\_flg} == 6.  We elect not to extend the analysis to sources with \texttt{var\_flg} == 5 because doing so would increase our storage and computational overhead by $\sim$60\% and we expect that a low proportion ($<$5\%) would be periodic.}  We only consider sources having \texttt{cc\_flags} == 0 to ensure that there are no imaging artifacts and \texttt{ext\_flg} $\leq$ 1 to ensure that it is not an extended source.  After applying \edit1{this selection}, we download $\approx$ 500,000 light curves from the AllWISE multi-epoch photometry table and the NEOWISE-R single exposure source table using a 1\arcsec matching distance.  

\edit1{Due to the $\sim$3 year gap between AllWISE and NEOWISE-R, it is possible that some high-proper motion AllWISE sources are not matched to their NEOWISE-R counterpart and are omitted from our analysis.  This could cause us to miss some interesting objects such as nearby, high-proper motion brown dwarfs.  To estimate the magnitude of this effect, we utilize the AllWISE/Gaia DR2 cross match of \citet{marr19}.  Out of the $\sim$200,000 cross matched objects with \texttt{var\_flg} 6 through 9 that satisfy the above selection, only $\sim$ 200 have a proper motion of above 100 mas/year so we estimate that $\sim$0.1\% of our variables suffer from a missed cross match due to high proper motion. For an analysis of high-proper motion sources in WISE see \citet{luhm14}.}

To eliminate redundancies and possible extraneous matches in the table, the \texttt{allwise\_cntr} listed in the NEOWISE-R single exposure table is mapped to the \texttt{source\_id\_mf} in the AllWISE multi-epoch photometry table.\footnote{\url{http://wise2.ipac.caltech.edu/docs/release/neowise/expsup/sec2_1a.html}} We exclude sources whose NEOWISE-R data is mapped to more than one AllWISE source. We also exclude AllWISE sources with no corresponding NEOWISE-R data because the number of AllWISE-only observations is not sufficient for our time-series analysis. 

For the AllWISE multi-epoch photometry, the following \edit1{selection} is applied: \texttt{saa\_sep} $>$ 5.0 deg (image outside of the South Atlantic anomaly), \texttt{moon\_masked} == 0000 (frame unaffected by light scattered off the moon),  and \texttt{qi\_fact} $>$ 0.9 (only the highest quality frames)\footnote{\url{http://wise2.ipac.caltech.edu/docs/release/allwise/expsup/sec3_1a.html}}.  In addition, points with \texttt{null} photometric measurement uncertainty or \texttt{null} for the reduced $\chi^2$ of the W1 profile-fit are excluded. For NEOWISE analogous cuts are applied\footnote{\url{http://wise2.ipac.caltech.edu/docs/release/neowise/expsup/sec2_3.html}} and we also exclude points with with \texttt{null} W1 profile fit signal-to-noise ratio.  The number of data points reported in our tables is the total remaining after these quality cuts are applied. Example python code for lightcurve download with quality flags is available through Github\footnote{\url{https://github.com/HC-Hwang/wise_light_curves}}.

\subsection{Time Series Analysis}

\begin{figure*}
\centering
\includegraphics[scale=1.0]{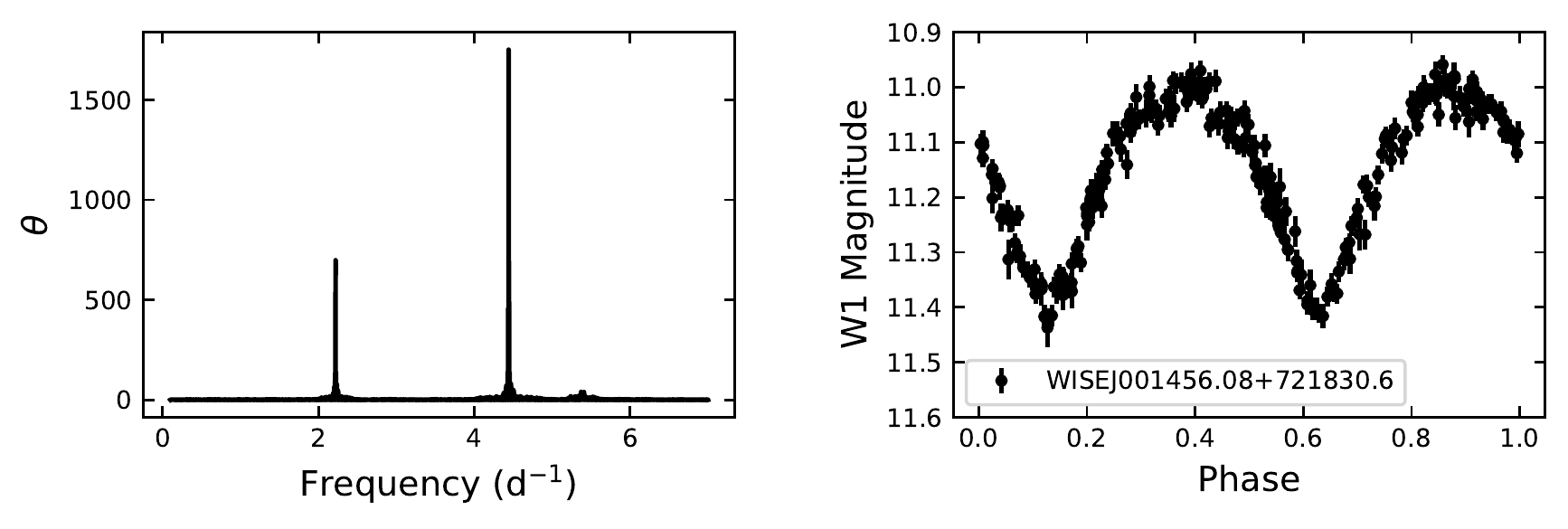}
\caption{\small {\bf Left:} An example periodogram made with WISE data. The vertical axis shows the value of the statistic, $\theta$, for a given frequency.  The higher the value of $\theta$, the lower the false alarm probability. The periodogram peak occurs at a frequency of $f_{max}$ $=4.44$  days$^{-1}$ \edit1{($P_{mhaov} = 0.225$ days)}. {\bf Right:} Phase-folded light curve of the same object. This object is an eclipsing binary with approximately symmetric eclipses.  This means that the \edit1{MHAOV period, $P_{mhaov}$} i.e. the period picked out by the MHAOV periodogram, is half the actual orbital period. We fold this light curve with a frequency of $0.5\times f_{max} = 2.22$ days$^{-1}$ which corresponds to an orbital period of \edit1{$2\times P_{mhaov} = 0.45$ days.}}
\label{fig:periodogram}
\end{figure*}

The periodogram is a time-series analysis tool that allows for the location and characterization of periodic signals. For this paper, we use the multi-harmonic analysis of variance (MHAOV) periodogram \citep{schw96}, which has been shown to be high-performing in comparison to other algorithms \citep{grah13b}.  We fit the data with periodic, orthogonal functions and use a statistic, $\theta$, which is the ratio of the squared norm of the model over the squared norm of the residuals \citep{schw03}, to quantify the quality of the fit \citep{schw98}. The fitting procedure is carried out for a grid of test frequencies and a periodogram shows the dependence of the statistic value on the test frequencies. Figure \ref{fig:periodogram} shows an example periodogram in frequency space with a corresponding phase-folded light curve. 

For three model parameters, the MHAOV periodogram is statistically equivalent to the classic Lomb-Scargle periodogram \citep{lomb76, ferr81, scar82, schw99}.  We use 5 model parameters to allow for better sensitivity to anharmonic oscillations.  This corresponds to fitting the data in real space with with a Fourier series of 2 harmonics (\edit1{fundamental plus one additional harmonic} -- \citealt{schw99, schw03, lach06, grah13b}). 

We adopt a frequency grid spacing of 0.0001 days$^{-1}$ \citep{grah13b}.  We search the frequency range 0.1 to 20 days$^{-1}$.  \edit1{This lower bound is selected because, due to the WISE observing cadence and our requirement that sources have good phase coverage, we expect low sensitivity to long-period variables.}  With regards to the upper bound, this extension beyond the traditional Nyquist limit of $\sim$7.6 days$^{-1}$ that corresponds to uniform sampling at the WISE orbital cadence is justified when the data is not uniformly spaced (see subsection \ref{subsec:per_uncert} for further discussion).

\begin{figure*}
\centering
\includegraphics[scale=1.0]{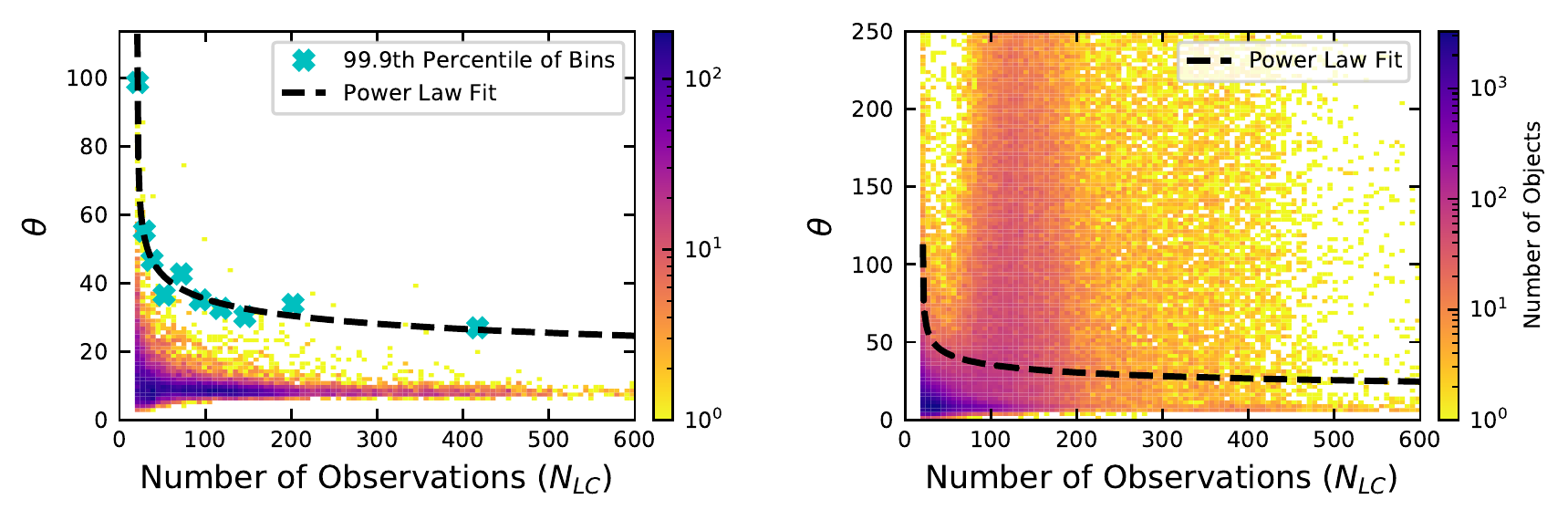}
\caption{{\bf Left:} Distribution of maximum statistic values versus the number of observations in the light curve ($N_{LC}$) for periodograms calculated on \edit1{20,000} white-noise light curves.  The data are binned based on $N_{LC}$.  The smallest bin contains about \edit1{1,300 light curves and most bins have about 2,000}.  Cyan crosses show an estimate of the 99.9$^{th}$ percentile of the corresponding statistic values in each bin.  The black dashed line shows the best-fit power law model using LMFIT \citep{newv14}. {\bf Right:} Maximum statistic value \edit1{from the periodogram} versus $N_{LC}$ for all WISE light curves with more than 17 data points and no peak at the WISE observing cadence.  The black dashed line represents the power law fit from the left panel.  For sources above this black line, we reject the null hypothesis that the observed light curve is consistent with white noise.} 
\label{fig:sig_sim}
\end{figure*}

\subsection{Periodogram Peak Significance}

\edit1{All the light curves in our sample have been determined by \citet{hoff12} to exhibit variations in brightness.  We seek to determine whether these variations are periodic in nature.}  When we run a periodogram on a source and observe a peak, we seek to reject the null hypothesis that the \edit1{variation displayed by the} light curve in question is pure noise and that the highest peak in the periodogram results from the chance alignment of random errors.  Although some inroads have been made toward an analytical understanding of peak significance (e.g. \citealt{horn86, koen90, schw99}), the results depend in complex ways on the nature of the data and the chosen test frequencies.

We adapt the Monte Carlo method of \citet{fres07, fres08} to quantify peak significance.  We generate \edit1{20,000} white-noise light curves that have means, flux deviations from mean, observing cadences, and individual data point photometric uncertainties that are representative of our actual sample.  \edit1{These light curves are variable at a level similar to that in the observed data, but contain no periodic signal by construction.}  To generate these light curves, we start with \edit1{20,000} randomly-selected WISE light curves.  For each light curve, there is an array of time measurements, $t_{i}$, an array of magnitude measurements, $m_{i}$, and an array of photometric uncertainties on each data point, $\sigma(m_i)$ where $i$ ranges from 1 to the number of data points in the light curve, $N_{LC}$.  For each light curve, we calculate the mean of the magnitude distribution, $\langle m \rangle$, and define a measure of its width, $w$, as the difference between the 84.2$^{th}$ and 15.8$^{th}$ percentile \edit1{(the percentiles marking, respectively, one standard deviation above and below the mean for a Gaussian distribution)}.  Next, we create a Gaussian distribution with mean $\langle m \rangle$ and standard deviation $0.5w$ and randomly draw $N_{LC}$ values from this distribution to create a new array of Gaussian (white) noise data, $noise_i$.  By substituting $noise_i$ for $m_i$ for each light curve, we create \edit1{20,000} white-noise light curves.  

We run a MHAOV periodogram on each of the white-noise light curves using the same frequency grid and number of model parameters that were used on the actual data.  \edit1{The left-hand panel of} Figure \ref{fig:sig_sim} shows the resultant distribution of maximum statistic values from the periodogram ($\theta$), as a function of the number of observations in the light curve ($N_{LC}$).  When $N_{LC}$ is low, the light curve contains less information, it is harder to distinguish periodic and aperiodic signals, and it is expected that a higher statistic value is required to credibly reject the null hypothesis.  To make \edit1{the left-hand panel} of Figure \ref{fig:sig_sim} we bin the data based on $N_{LC}$.  Every bin has \edit1{$\sim$2,000 simulated light curves save the last one on the right which has $\sim$1,300}.  In each bin, we study the resultant empirical cumulative distribution function of the calculated maximum statistic values and estimate the $99.9^{th}$ percentile. Then we fit a power law model of the form $\theta_{cutoff}(N_{LC}) = A(N_{LC}-N_0)^k$ to estimate the appropriate cut-off value as a function of $N_{LC}$ to reject the null hypothesis.  \edit1{The best-fit values are $A=79 \pm 8$, $N_0 = 20.7 \pm 0.2$, and $k=-0.18 \pm 0.03$}.  For sources that lie above this best-fit line we reject the null hypothesis that the peak results from the chance-alignment of random errors for a white-noise light curve \edit1{(see Figure \ref{fig:sig_sim} right)}. \edit1{For $N_{LC} \leq 20$ the above fit diverges.  Due to other cuts detailed below in Subsection \ref{subsec:selection}, the minimum number of points that are required for periodic variable detection is 18.  For sources with between 18 and 20 data points, we require a safe cut of 125 on maximum statistic.  This only ends up adding one additional periodic source to the catalog.}

\subsection{Completeness} \label{subsec:complete}

The above cut on the maximum statistic seeks to limit the false alarm probability; however, it does not say much about the probability of periodicity itself nor the incidence of periodic signals that are passed over \citep{vand18}. \edit1{The completeness, that is, the proportion of truly periodic signals we expect to detect, is a complex function of the nature of periodicity \citep{schw99}, the observing cadence, and the data quality and quantity.}  It depends on the magnitude (worse for very faint and very bright objects), amplitude (worse for small amplitude), signal shape and phase (worse for short-duration pulses, especially if they occur in between observing epochs), period (worse for extremely short and long periods), and number of observations (worse for fewer observations). 

\begin{figure}
	\includegraphics[scale=1.0]{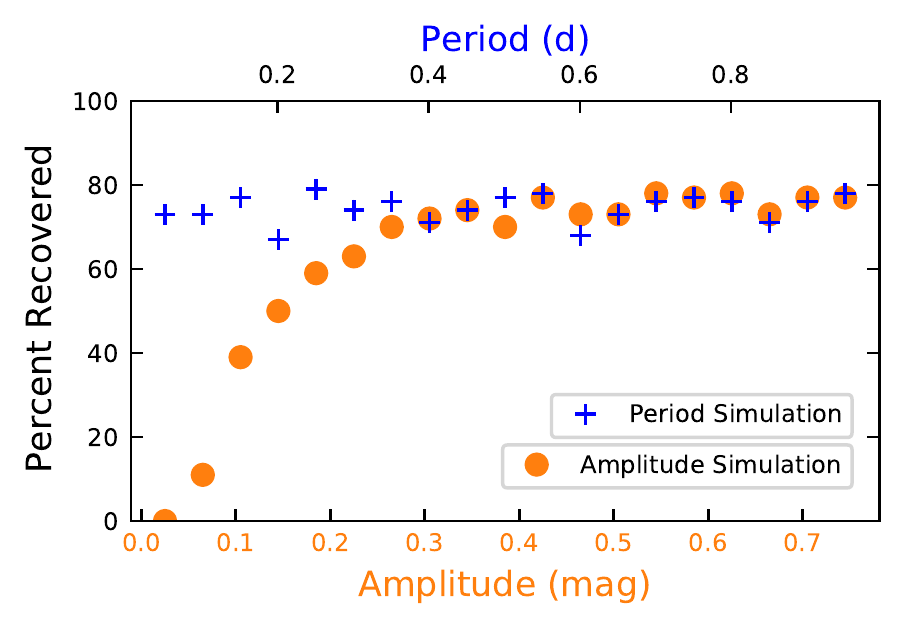}
	\caption{\edit1{Percent of signals correctly identified as periodic after running our method on a random sample of WISE light curves with a simulated sinusoidal signal.  For the period simulation, the amplitude was fixed at a value representative of our close binaries.  For the amplitude simulation, the period was fixed at a value representative of our close binaries.}}
	\label{fig:completeness}
\end{figure}

Completeness is not the main focus of this project, and we do not attempt to characterize completeness across all of these parameters.  Instead, \edit1{we briefly consider just the special case of contact and near-contact binaries with sinusoidal and near-sinusoidal light curves.}  We explore how our ability to detect such sources varies as a function of signal amplitude and period.  We randomly select 100 variable WISE light curves that have \edit1{8$<$W1$<$15 mag}.  We then simulate a sinusoidal signal centered at the W1 magnitude of the light curve and sample it at the original cadence, preserving the individual data point uncertainty associated with each timestamp.  In the first simulation, we fix the phase and choose an amplitude characteristic of our recovered periodic variables (0.4 mag).  We pick a starting period of 0.051 days and calculate the MHAOV periodogram, phase-fold the result and see if the source would have been classified as periodic (using the criteria detailed below in Subsection \ref{subsec:selection}).  We iterate over all light curves for a given period and then increase the period and repeat until the period exceeds a value of \edit1{1 day}.  We choose periods between 0.05 and 1 day because the majority of our periodic sources are contained in this range.  Next, we repeat the same procedure but this time fixing the period at a value of 0.22 days (a value typical of our close binaries) and varying the amplitude between value of \edit1{0.025 and 0.75} magnitude. \edit1{Figure \ref{fig:completeness} shows the results of these simulations.}  In summary, given our data and our method, we estimate that we can detect $\sim$75\% of near-sinusoidal signals with with periods between 0.05 and 1 day and peak-to-peak amplitudes above 0.25 magnitude. \edit1{As mentioned above, completeness is a function of signal shape, so these conclusions will not necessarily hold for strongly non-sinusoidal signals.  We elect to focus solely on sinusoidal signals because contact and near-contact EBs dominate our sample and many of their light curves are well-approximated by a sinusoid.}

\subsection{Period Uncertainty} \label{subsec:per_uncert}

\edit1{After determining that a periodogram has one or more significant peaks and represents a periodic variable candidate, we next determine whether it is indeed a truly periodic signal.  If multiple periodogram peaks are present, it is important to determine which among them, if any, corresponds to the correct period.}  The observed light curve is a product of the continuous underlying signal and the discrete and unevenly-spaced window function (i.e. observational cadence).  Aliasing, or the correlation between frequencies equidistant from one-half of the the inverse sampling rate (Nyquist frequency) is familiar in the case of evenly-spaced sampling.  Uniform observations at the WISE satellite period ($\sim 95$ minutes) would correspond to a Nyquist frequency of $\sim 7.6$ days$^{-1}$. Deviations from uniformity dampen the effects of aliasing and allow for the detection of periodic components above the traditional Nyquist limit \citep{eyer99, koen06}. 

Despite the orthogonality of the multi-harmonic periodogram fit at each individual frequency, the fits on any set of frequencies are generally not independent in the case of irregular sampling \citep{schw98}. This means that the statistic values at different frequencies can be correlated.  Sometimes these correlations manifest as separate peaks. Compared to other algorithms, we find the MHAOV periodogram to be effective at damping secondary peaks across the range of probed frequencies.  For the purposes of this analysis, we use the \edit1{MHAOV period ($P_{mhaov}$)} as given by the periodogram.  In the case of \edit1{light curves with two identical or near-identical minima per cycle}, such as an eclipsing binary composed of similar stars, \edit1{$P_{mhaov}$} corresponds to half of the orbital period.  

Correlations also cause the periodogram peak to have finite width and limit the precision with which the frequency can be determined from the peak.  Generally, the uncertainty is a function of signal characteristics, the model used, the temporal baseline, and the quantity and quality of data (see \citealt{hart08, lach09, hard13} for examples of different error estimation strategies).

To refine our frequency measurements and to estimate their error, we repeat the periodogram procedure with a refined frequency grid in the vicinity of the main periodogram peak.  Our frequency measurement is the position of the likelihood peak on the refined frequency grid. We use the width of the peak on the fine grid and the surrounding background noise level to estimate the frequency error $\sigma_{freq}$ \citep{schw91, schw95, schw96}.

\begin{figure}
    \centering
	\includegraphics[scale=1.0]{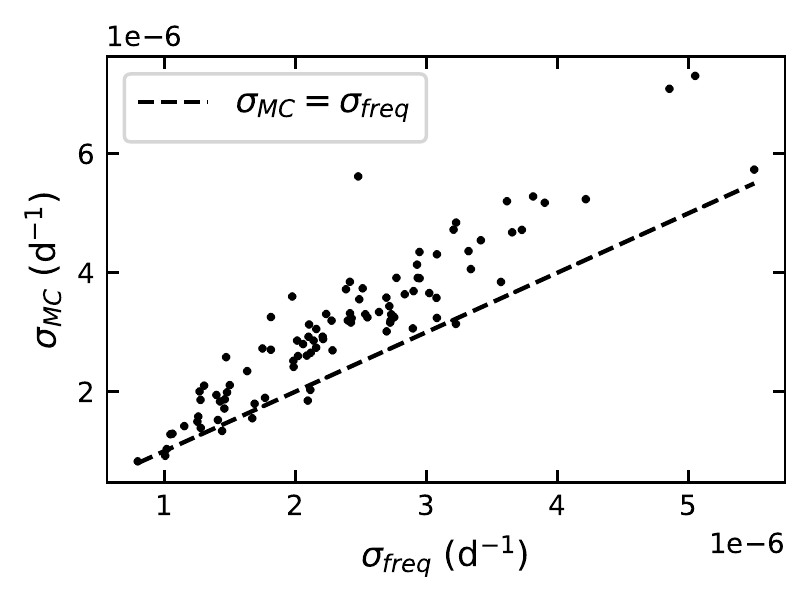}
	\caption{Frequency error derived via MCMC simulations, $\sigma_{MC}$, versus the error derived from periodogram methods, $\sigma_{freq}$, for a set of randomly selected periodic light curves. The dashed line is a plot of $\sigma_{MC} = \sigma_{freq}$. The two error estimates are highly correlated and the periodogram-based error estimate is systematically smaller than the MC-based error estimate.}
	\label{fig:mc_compare}
\end{figure}

\begin{figure*}
    \centering
	\includegraphics[scale=1.0]{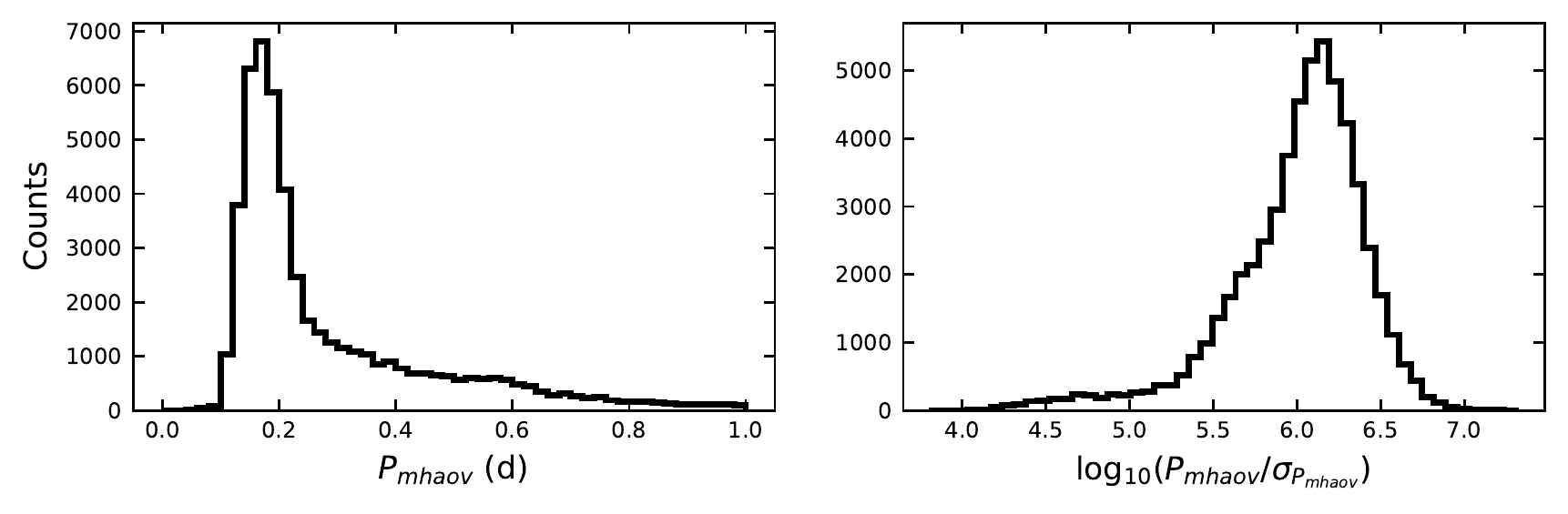}
	\caption{{\bf Left: } Distribution of \edit1{MHAOV (i.e. as calculated by the MHAOV periodogram) periods \edit1{($P_{mhaov}$)}} for 50,493 periodic variables with MHAOV periods in the range 0.05 to 1 $d$.
	{\bf Right: } Distribution of \edit1{$\frac{P_{mhaov}}{\sigma_{P_{mhaov}}}$}
	for the 56,177 identified periodic variables}.
	\label{fig:p_over_sigp}
\end{figure*}

To check this frequency error estimation scheme, we compare with errors derived using \texttt{emcee} \citep{fore13}.  To have realistic errors for realistic (non-sinusoidal) light curves, we take a random sample of 100 light curves, phase-fold them, and fit them with a Fourier series using LMFIT \citep{newv14}.  Then, holding the Fourier coefficients fixed, we fit the light curve in the time domain with \texttt{emcee} using the phase, frequency, and constant offset from zero as parameters.  The fitting is done in this way because the WISE cadence made simultaneously fitting the Fourier coefficients, phase, frequency, and constant offset intractable.  To get the frequency distribution, we marginalize over the constant and the phase. As seen in Figure \ref{fig:mc_compare}, the MC-derived errors,  $\sigma_{MC}$, are correlated with the periodogram-derived errors, $\sigma_{freq}$, but are systematically higher by $\sim$ 25\%.  In what follows we use a conservative error estimate by multiplying the periodogram-derived error by a factor of 1.25.  The period errors are on the order of $\sim10^{-6}$ - $10^{-7}$ days, but they do not capture the error of picking the wrong periodogram peak entirely and thus should be used with caution \citep{vand18}.

\subsection{Selection of Periodic Variables} \label{subsec:selection}

To select periodic sources, we first require that the maximum statistic of the MHAOV periodogram, $\theta_{max}$, exceeds the cutoff value given by the power-law fit of Figure \ref{fig:sig_sim} for the appropriate number of points.  Our cuts on false alarm probability are less strict than those of some other surveys that identify periodicity.  This offers us the possibility of capturing sources undergoing interesting orbital evolution.  For a higher confidence sample, a more restrictive cut on the maximum statistic value should be used. In addition to the cut on the maximum statistic value, we require phase coverage of at least 90\%.  We estimate phase coverage by dividing the phase-folded light curve into 20 equally sized bins and then calculating the percentage of bins that contain at least 1 data point.  The requirement that phase coverage is at least 90\% implicitly requires that the sources have at least 18 data points.  Next, we exclude sources whose highest periodogram peak is at 1 or 0.5 times the orbital frequency of the WISE satellite ($f_{peak} \notin [7.52, 7.68]$ days$^{-1}$ and $f_{peak} \notin [14.65, 15.70]$ days$^{-1}$).  \edit1{Finally, we include a cut to limit phase offset due to period errors over the course of the time series.  We calculate phase offset via
\[\mbox{Phase Offset} =  \frac{\sigma_{P_{mhaov}}\times \rm{Baseline}}{(P_{mhaov})^2}\]
where the baseline is the time series duration, $\sigma_{P_{mhaov}}$ is the period error, and $P_{mhaov}$ is the period as given by the periodogram.  The phase offset can be thought of as the ratio of the period uncertainty to the period multiplied by the complete number of cycles that the signal undergoes over the time-series duration.  We require that the phase offset be less than 0.2.}  With the above cuts, we identify 56,177 periodic variable candidates.  Figure \ref{fig:p_over_sigp} shows the distribution of \edit1{MHAOV periods and MHAOV periods over the period errors for these sources.}

\section{Catalog Contents}\label{sec:cont}

\subsection{Catalog Contents and Gaia Cross-Match} \label{subsec:gaia}

We cross-match our periodic variable candidates with Gaia DR2 \citep{prus16, gaia18a, evan18, aren18} using the pre-computed, best-neighbor WISE/Gaia cross-match catalog of \citet{marr19}. We find 188,043 matches out of a total of 454,103 variable objects ($\sim 41.4 \%$) which is similar to the total percentage of AllWISE sources that have a best-neighbor cross-match (39.83\% - see \citealt{marr19}).  Of our 56,177 periodic variable candidates, 49,384 or $\sim 87.9\%$ have a best-neighbor Gaia cross-match.  The higher match rate for periodic variables is due to the fact that the aperiodic variables -- many of them stochastically varying young stellar objects -- tend to be redder, with a median $\langle$W1-W3$\rangle = 0.909$ mag as compared to the periodic variables with $\langle$W1-W3$\rangle = 0.311$ mag.  As a further check, we cross-match our sample with the WISE young stellar object catalogs of \citet{mart16} and find 44,196 matches of which only 237 are flagged as periodic.

We use the Gaia cross-match and considerations of the limitations of WISE to get a sense of the catalog contents and make the case that the sample is dominated by eclipsing binaries.  We search for periods between 0.05 and 10 days, but due to the WISE observing cadence \edit1{and our requirement of good phase coverage}, the sensitivity drops for periods above 2 days.  As a result, we do not expect many long-period variables in the more luminous part of the color-magnitude diagram.  The period sensitivity limit combined with the crowding in the Galactic disk cause us to detect few classical Cepheids.  \edit1{We expect to detect few $\delta$ Scuti variables because many have variability amplitudes that are too low to be reliably detected \citep{murp19} and many high-amplitude $\delta$ Scuti are faint and blue.} The photometric sensitivity of WISE also limits our ability to detect variability on the white dwarf sequence.

For periodic objects with a Gaia cross-match, in order to have robust absolute magnitudes and colors, we require:
\begin{enumerate}
    \item \texttt{parallax\_over\_error} $>$ 10
    \item \texttt{phot\_g\_mean\_flux\_over\_error} $>$ 50
    \item \texttt{phot\_bp\_mean\_flux\_over\_error} $>$ 10
    \item \texttt{visibility\_periods\_used} $>$ 8
    \item \texttt{phot\_rp\_mean\_flux\_over\_error} $>$ 10
\end{enumerate}

In addition, we restrict \texttt{phot\_bp\_rp\_excess\_factor} in accordance with \citet{gaia18b}.   After applying these cuts we retain 34,857 sources.

\begin{figure}
	\includegraphics[scale=1.0]{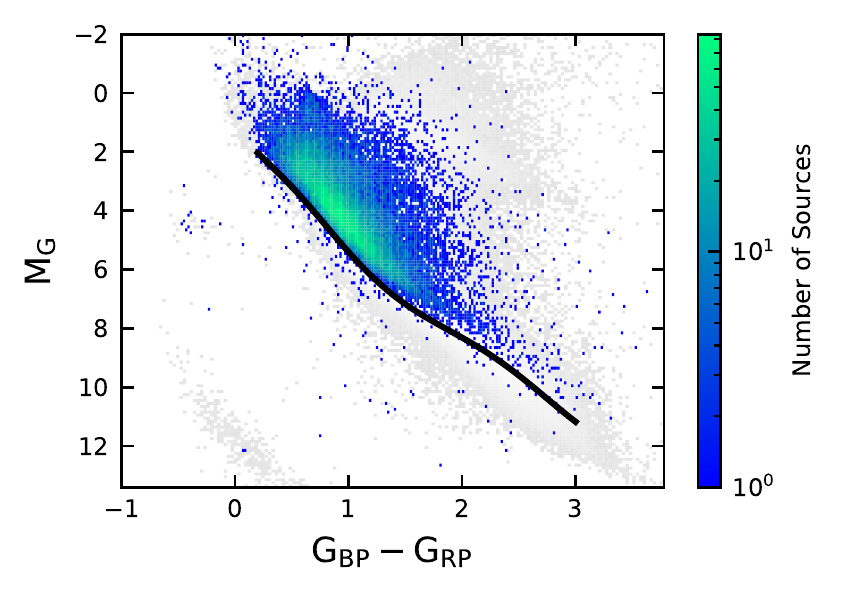}
	\caption{\edit1{Our periodic variables cross-matched with Gaia and shown on the color-absolute magnitude diagram.  The black line is the Pleiades cluster spline fit from \citet{hame19}.  The gray background sources are a sample of 500,000 Gaia sources chosen in accordance with \citet{gaia18b} that are shown for reference.}}
	\label{fig:HR_ms_spine}
\end{figure}

\edit1{Figure \ref{fig:HR_ms_spine} shows the Gaia color-absolute magnitude diagram for all all the WISE periodic variables with a Gaia cross match that satisfy the above quality cuts.  The Pleiades main-sequence fit of \citet{hame19} is shown in black. Most of the WISE periodic variables are main-sequence stars. On the main sequence, the most common form of periodic stellar variability is eclipsing binaries.  These sources are located above the main-sequence fit because there is extra flux from the presence of the second star.   A few periodic sources are located below the main sequence. They may be white dwarf-brown dwarf binaries or cataclysmic variables.  The gray scale background is a sample of 500,000 Gaia sources that meet the selection of \citet{gaia18b} that is shown for reference.}

\begin{figure}
	\includegraphics[scale=1.0]{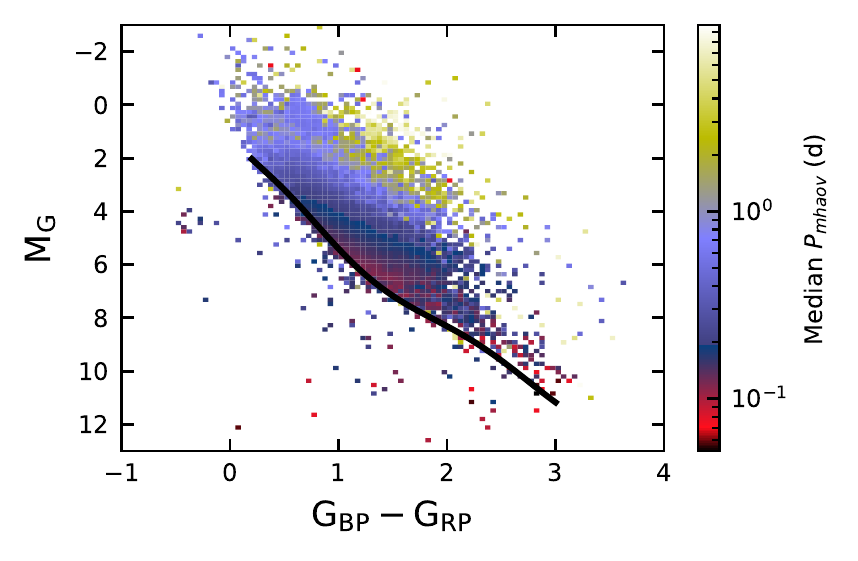}
	\caption{Color-absolute magnitude diagram of our WISE periodic variables color-coded by the median \edit1{MHAOV period} in each color-absolute magnitude bin. The main-sequence fit of \citet{hame19} is shown in black for reference.}
	\label{fig:CMD_color_code}
\end{figure}

Figure \ref{fig:CMD_color_code} again shows the Gaia color-absolute magnitude for the same sources color-coded by the median \edit1{MHAOV period} in each color-absolute magnitude bin.  For our sample, on average, more massive sources tend to have longer periods.  For close binaries, this makes sense because larger stars have a longer limiting period before reaching contact.  A group of pulsating RR Lyrae stars can be seen at $M_G\sim1$ and BP-RP$\sim0.7$ with periods between $\sim$0.2 days and 1 day \citep{pres64, kole12, das18}. These sources are in purple and are distinguished from the green band that represents the surrounding EBs in the color-magnitude diagram.   

\begin{figure}
\includegraphics[scale=1.0]{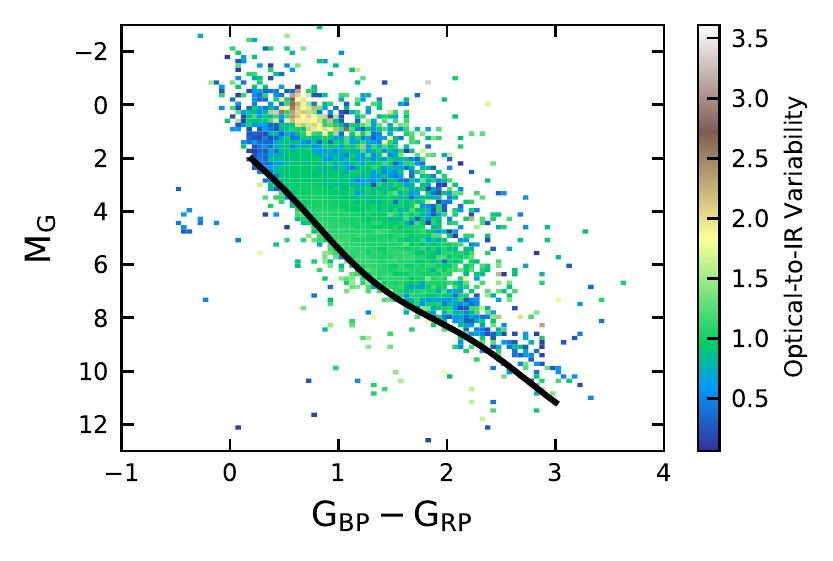}
\caption{Gaia color-magnitude diagram color-coded by the median ratio of optical fractional variability from Gaia G band to infrared (IR) fractional variability from WISE W1 in each color-absolute magnitude bin.  Gaia fractional variability is calculated in accordance with \citet{hwan20}.  \edit1{We plot only those sources that have a Gaia fractional variability of at least 1\% and have a fractional variability ratio of at least 0.05.}  A group of RR Lyrae can be seen in tan/brown that vary more in the optical than in the infrared.  The majority of the sources are eclipsing binaries and have equivalent optical and infrared variability amplitudes (i.e. Optical-to-IR ratio of $\sim$1). The black line is the Pleiades main-sequence fit of \citet{hame19}}
\label{fig:CMD_opt_IR}
\end{figure}

\edit1{Figure \ref{fig:CMD_opt_IR} shows another view of the Gaia color-absolute magnitude diagram color-coded by the ratio of fractional variability in the optical to fractional variability in the infrared (IR).  \edit2{To calculate the optical fractional flux variability from Gaia, $\sigma_{optical}$, we follow the methods of \citet{hwan20}.  We compute the fractional flux variability from the Gaia photometric errors and then account for instrumental errors.  We compute the IR fractional flux variability, $\sigma_{IR}$, by $\sigma_{IR} = 0.4\ln(10)*\textrm{Std}_{\textrm{W1}}$, where $\textrm{Std}_{\textrm{W1}}$ is the standard deviation of W1 in magnitudes, and the factor $0.4\ln(10)$ comes from the unit conversion between fluxes and magnitudes.  We plot only those sources here for which the ratio of optical-to-IR fractional variability is greater than 0.05 and which have a fractional variability of at least 1\% in Gaia. }  Most of the sources have approximately equal optical and IR fractional variability, consistent with what is expected for eclipsing binaries.  These sources are also offset to the bright side of the main-sequence fit indicating that there is excess flux due to the presence of a second star.  A group of RR Lyrae can be seen that vary more in the optical than in the infrared due to the nature of their pulsation \citep{das18}.}

\subsection{Comparison with \citet{chen18} Catalog} 

\citet{chen18} present a catalog of 50,282 periodic variables from WISE identified with the Lomb-Scargle \citep{lomb76, scar82} periodogram and classified via light curve fitting.  Our work differs from that of \citet{chen18} in that we extend the period search grid to shorter period objects, use a different periodogram, involve a Gaia DR2 cross-match, use data from a more recent NEOWISE release, and apply a different selection.  \citet{chen18} also discarded sources with varying periods from AllWISE to NEOWISE and we make no such restriction.

We cross-match our periodic variables with those of \citet{chen18} using a matching distance of 1 arcsecond.  Increasing the matching distance to 5 arcseconds does not significantly alter the results.  \edit1{A cross-match was necessary because \citet{chen18} used a truncated form of the WISE designation that made it difficult to use this for matching purposes.} Of all the sources with 6 $\leq$ \texttt{var\_flg} $\leq$ 9 that we analyzed, 41,532 have a cross-match with the 50,282 periodic variables of \citet{chen18}.  Their remaining 8,750 variables were excluded by our initial selection and were never a part of our periodogram analysis.  8,703 of the variables were excluded by our cuts on \texttt{cc\_flags} to avoid contamination and confusion and the remaining 47 were excluded by our cuts on \texttt{ext\_flg} to remove extended objects. Of the 41,532 variables in common, we mark 36,991 as periodic.  We miss some of the sources because our more stringent cuts on the quality of individual exposures reduces the number of points in the light curve and prevents us from detecting them as periodic.   \edit1{We explore several methods for comparing the periods from our two catalogs.  First, we consider the raw period difference. For 33,870 ($\sim$92\%) of the cross-matched periodic sources there is a period difference $\Delta P_{half} = |0.5P_{chen18} - P_{mhaov}| < 0.001$ d (relevant for binaries with 2 similar minima per cycle)  or $\Delta P = |P_{chen18} - P_{mhaov}| < 0.001$ d.  Second, we incorporate uncertainties to see if the period differences mentioned above are significantly different than zero.  We find that for more than 96\% of the sources, either $\Delta P_{half}$ or $\Delta P$ are within three standard deviations from zero.  Third, we explore a more stringent selection for compatible periods by requiring:
\edit1{
\[\left| \frac{P_{mhaov} - P_{chen18}}{P_{mhaov}}\right| \times Num\_Cycles < 0.20\]
or
\[\left| \frac{P_{mhaov} - 0.5P_{chen18}}{P_{mhaov}}\right| \times Num\_Cycles  < 0.20\].}}

\edit1{Here $Num\_Cycles$ is the total number of periodic cycles completed over the duration of the time series, $Num\_Cycles = Baseline/P_{mhaov}$.}
\edit1{We find that 27,568 ($\sim$75\%) of the cross-matched sources satisfy this third criterion.  For all three compatibility criteria, the incompatible periods generally correspond to cases in which either (a) our more stringent cuts reduce the number of data points and leads to a different period estimate, (b) the detected period is greater than 5 days i.e. in a range poorly probed by WISE, or (c) we measure a period outside the range probed by \citet{chen18} indicating that one of the measurements is a harmonic.  Of these effects, we expect (a) to be the most common.  For example, for objects with $\Delta P_{half}$ or $\Delta P$ more than three standard deviations from zero, on average, our periodogram is computed on $\sim$ 39 fewer data points (despite the fact that we include more years of NEOWISE data) due to our more stringent cuts on the quality of the individual exposures.  For objects with incompatible periods as determined by the comparison method that takes into account the number of cycles, our post-quality-cut light curves had, on average, 22 fewer data points than the light curves of \citet{chen18}.  Due to the long observing baseline ($\sim$3100 d), even a slight change in the single-exposure selection criteria can lead to a period shift large enough to cause the source to be classified as incompatible under this third comparison method.}

For $\sim$89\% of the matches, our \edit1{MHAOV periods, $P_{mhaov}$,} are one-half the period cited by \citet{chen18}, $P_{chen18}$, which is expected for a sample dominated by close eclipsing binaries where the primary and secondary eclipses are difficult to distinguish.  Over 98\% of the sources for which we recover half periods have $\Delta P_{half} < 0.0001$ days. \citet{chen18} classify 32,151 of the cross-matched sources as binaries, 710 as some type of Cepheid, and 2,319 as RR Lyrae.  The remainder were assigned an ambiguous classification (Cepheid/Binary, Binary/RR Lyrae, or miscellaneous) due to a lack of characteristic features in their infrared light curves. 

Finally, 19,186 of our periodic variables were not in the periodic variable catalog of \citet{chen18} because of our different period search method and our expanded period search grid on the short-period end. 
\begin{center}
\begin{table*}[h!]
\begin{center}
\begin{tabular}{llll}
\hline\hline
Column & Data Type & Units & Description \\
\hline
wise\_id & str19 & -- & WISE designation \\
ra & float64 & deg & Right ascension \\
dec & float64 & deg & Declination \\
sigra & float64 & arcsec &  Right ascension error \\
sigdec & float64 & arcsec &  Declination error \\
w1mpro & float64 & mag &  W1 magnitude \\
w1sigmpro & float64 & mag &  W1 magnitude error \\
w1snr & float64 & mag &  W1 signal-to-noise ratio \\
w2mpro & float64 & mag &  W2 magnitude \\
w2sigmpro & float64 & mag &  W2 magnitude error \\
w3mpro & float64 & mag &  W3 magnitude \\
w3sigmpro & float64 & mag &  W3 magnitude error \\
w4mpro & float64 & mag &  W4 magnitude \\
w4sigmpro & float64 & mag &  W4 magnitude error \\
var\_flg & bytes4 & -- &  Variability flags for all four bands \\
num\_pts & float64 & -- &  Number of observations in light curve after quality cuts  \\
median\_err & float64 & mag & Median individual data point uncertainty for light curve \\
mean\_mag & float64 & mag &  Mean magnitude \\
std\_mag & float64 & mag &  Standard deviation of magnitude \\
median\_mag & float64 & mag & Median magnitude\\
amp & float64 & mag &  Amplitude \\
FF & float64 & -- &  Fainter Fraction \\
rel\_asym & float64 & -- & \edit1{Relative Asymmetry \citep{zaka14}} \\
M & float64 & -- & \edit1{Magnitude Ratio/M-test value \citep{kine06}} \\
skew & float64 & -- &  Skewness \\
kur & float64 & -- &  Kurtosis \\
phase\_cov & float64 & -- &  Phase coverage \\
min\_mag & float64 & mag & \edit1{Minimum magnitude} \\
max\_mag & float64 & mag & \edit1{Maximum magnitude} \\
baseline & float64 & d & \edit1{Time series duration} \\
R & float64 & -- &  Ratio of variability amplitude on short timescales to that on long timescales \\
max\_stat & float64 & -- &  Maximum statistic value \\
cutoff\_stat & float64 & -- &  Cutoff maximum statistic value to reject null hypothesis \\
periodic & bool & -- &  Periodic sources receive a value of True \\
P\_mhaov & float64 & d &  \edit1{MHAOV period} \\
sigP\_mhaov & float64 & d &  \edit1{MHAOV period error} \\
EB & bool & -- &  Eclipsing Binaries receive a value of True \\
\hline
\hline\\
\end{tabular}
\end{center}
\caption{Column descriptions for our catalog of WISE variables.  \edit1{After careful consideration, we elect to report the MHAOV period ($P_{mhaov}$) i.e. the period as given by the periodogram.  Based on our cross match with \citet{chen18}, we estimate that for more than $\sim$ 90\% of the EBs, $P_{mhaov}$ corresponds to one-half of the orbital period.  However, without more detailed light curve fitting or analysis with a different periodogram (e.g. the string-length methods of \citealt{lafl65, burk70, rens78, dwor83}) which is outside the scope of this work, there is no clear way to identify those binaries for which we detect the actual orbital period and not a harmonic.  It depends on the data quality/cadence and the particular signal shape in question.  When no measurement could be made for a given source and column, the value is missing.  Not all non-parametric measures are calculated if the source had fewer than 5 data points. }  Each band has its own variability flag.  If the variability flag for a band is `n', then no flag was assigned in that band.  The complete catalog is available \href{https://zakamska.johnshopkins.edu/data.htm}{online}}
\label{table:data_model}
\end{table*}
\end{center}

\section{Non-Parametric Methods for Light Curve Analysis}\label{sec:nonparm}

\edit1{We present a catalog of 454,103 WISE variables with a variety of non-parametric measures and periodogram measurements.  We explicitly flag periodic variables and eclipsing binaries.  The data model is shown in Table \ref{table:data_model}.  The non-parametric measures and the identification of eclipsing binaries are the subjects of this section.}

\subsection{Non-parametric measures of variability}

\edit1{In this section, we explore non-parametric and computationally-inexpensive methods for light curve analysis.}  The distribution of observed fluxes, without reference to their time dependence, carries a wealth of information.  If the cadence is suitable, the distribution of the observed fluxes can be assumed to be randomly drawn from (and therefore to be representative of) the underlying flux probability density function (PDF). The various moments of the observed flux distribution, assumed to be representative of the moments of the underlying PDF, are easy to measure and computationally fast.  \edit1{Various measures of these moments with different weighting schemes have been successfully utilized for variable star classification in a variety of other works (e.g. \citealt{kine06, duba11, rimo12, rimo13, drak13, bass16, rimo19}).}

The first and the second moment -- related to the mean magnitude and the r.m.s. variations around the mean -- are already directly or indirectly incorporated into our analysis. In particular the second moment of the PDF is related to the \texttt{var\_flg} we use as the primary step in our selection. We also compute a non-parametric amplitude (henceforth \emph{amplitude}), which is the range between the 5th and 95th percentile of the magnitude values.

The third moment (skewness) of the PDF can help distinguish between eclipsing and eruptive types of variability. For an object which stays at constant magnitude and occasionally undergoes dimmings due to occultations or eclipses, the PDF is expected to have a tail to fainter magnitudes (which correspond to the fluxes during occultations). For an object that instead undergoes occasional eruptions, we expect a tail to brighter magnitudes. \edit1{In addition to the standard Fisher-Pearson skewness, an assortment of other non-parametric measures of the the third moment have been introduced.  One such measure is the magnitude ratio or M test of \citet{kine06}.  This is calculated as 
\[M = \frac{max - median}{max - min}\]
where max, median, and min refer to the maximum, median, and minimum of the magnitude distribution respectively.  Another such measure is the relative asymmetry (\emph{RelAsym}), introduced by \citet{zaka14} in the context of velocity profiles but also relevant to the analysis of time series data.  This is calculated by
\[RelAsym = \frac{(P_{95} - P_{50}) - (P_{50} - P_{5})}{(P_{95} - P_{5})}\]
where $P_{95}$, $P_{50}$, and $P_{5}$ refer to the $95^{th}$, $50^{th}$, and $5^{th}$ percentiles of the magnitude distribution respectively.}

\begin{figure*}
\centering
\includegraphics[scale=1.0]{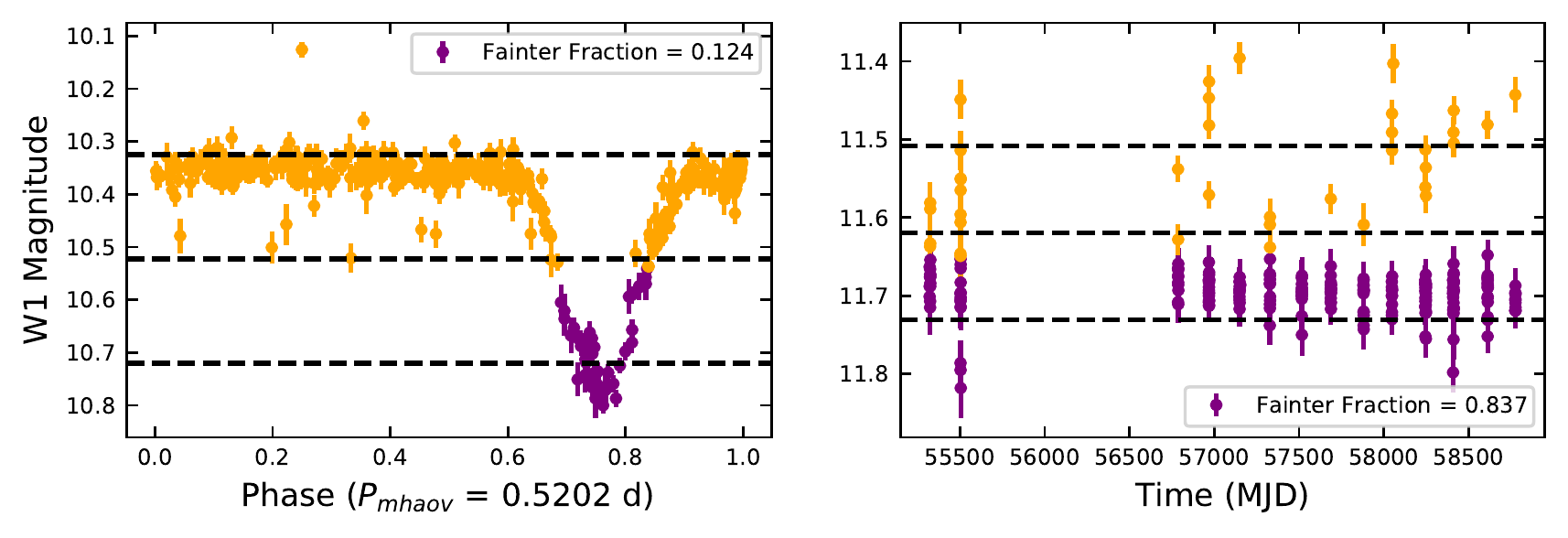}
\caption{{\bf Left:} The light curve of a an Algol-type eclipsing binary (WISEJ234343.55+812751.6) phase-folded with the \edit1{MHAOV period ($P_{mhaov}$) as given by the periodogram}, thus showing only one eclipse instead of two. Projecting the light curve on the brightness axis gives us the flux probability density function (PDF), which can be used for non-parametric measures of the light curve.  The top and bottom black, dashed lines show the 5th and 95th percentiles of the PDF.  The difference between these two lines is an estimate of the signal amplitude.  The middle line represents the halfway point between these two. The points in purple are the \emph{fainter fraction} of the light curve. Stars with eclipses and stars with eruptions show different PDFs and can be distinguished based on these metrics. {\bf Right:} Similar idea but for an eruptive variable (WISEJ005714.34-703745.5).  Eruptive variables are typically aperodic so we display the actual light curve instead of the phase-folded one.}
\label{fig:nonpar}
\end{figure*}

In this paper, we introduce a new non-parametric measure, \emph{fainter fraction} (\emph{FF}), which is related to the third moment. Specifically, for each light curve, we calculate the halfway point between the $5^{th}$ and $95^{th}$ percentile of magnitude and then calculate the fraction of points with magnitudes greater (fainter) than this halfway point by at least their measurement uncertainty.  If \emph{FF} is greater than 0.5, this means that the object spends most of its time in the faint state, becoming brighter less than half of the time, and it is likely to be of an eruptive type. The use of this measure is demonstrated in Figure \ref{fig:nonpar}. 

\edit1{Each of the aforementioned measures has its own advantages and we report all of them (i.e. skewness, magnitude ratio, \emph{RelAsym}, and \emph{FF}) in our catalog for all of the WISE variables so that users can choose which to employ.  Due to the fact that we aim to apply these measures to all of the variables, periodic and not, we do not employ the moment weighting scheme proposed by \citet{rimo13} because this would require a linear interpolation in phase (interpolation in time is not possible for the WISE cadence) which is not possible if the source is aperiodic.  The correlations between each of these measures are shown in Table \ref{table:corr_matrix}. All these measures are dimensionless, but the standard skewness is not bounded like \emph{M}, \emph{FF}, and \emph{RelAsym} making it less desirable for use with a classifier.} 

\edit1{For this work, we utilize our new measure, \emph{FF} because we find that it performs well in comparison with the these other measure for our particular use case.  The magnitude ratio, \emph{M}, is susceptible to effects from outliers and erroneous measurements due to its use reliance the maximum and minimum magnitude. Similar to \emph{RelAsym}, \emph{FF} is not as strongly affected by these outliers.  Another advantage of \emph{FF} is that, unlike both \emph{M} and \emph{RelAsym}, it is not normalized by some measure of the signal amplitude.  Using \emph{FF} allows us to disentangle the signal amplitude from the estimate of the third moment.  This is useful, for example, in the case of eclipsing binaries with narrow eclipses as these signals should be more localized in \emph{FF} space than in the space of \emph{RelAsym} or \emph{M}.  Furthermore, in Subsection \ref{subsec:find_ebs}, we find that the separation between RR Lyrae and EBs is cleaner in the \emph{FF}-\emph{amplitude} space than if \emph{FF} is substituted for one of these other measures.}

Finally, we introduce another non-parametric measure to describe the timescale of variability.  Due to the peculiar cadence of WISE observations -- \edit1{every $\sim$six months there is an bundle of $\sim$10 observations taken $\sim$95 minutes apart} -- WISE has sensitivity to variability on a ~1 day time scale, as well as to long-term variations.  We use $R$, the ratio of the average r.m.s. variability \edit1{on day/hour-long timescales} to the r.m.s. variability of the entire light curve. \edit1{To calculate the variability on short timescales, we remove long-term variations from the light curve.  We subtract off the mean from each observing bundle so that each bundle is centered at zero magnitude.  Then, we find the standard deviation of the light curve composed of the zeroed bundles.}  Given that WISE has only a limited number of $\sim$day-long visits to the same position on the sky, this method is computationally inexpensive. For sources with equal signal-to-noise (S/N) of variability, the ratio $R$ is expected to decline from 1 to 0 as we go from objects that vary on day-long timescales to those with only month- or year-long variability. Sources with low S/N of variability have both their short-term and long-term variability in line with the photometric $\sigma$.  \edit1{The idea behind this non-parametric measure is similar to that behind the \texttt{MEDIAN\_RANGE\_HALFDAY\_TO\_ALL} measure used by Gaia \citep[section 7.3.3][]{eyer18}, but instead of a sliding window, the one-day intervals for computation of short-timescale variability are naturally provided by the WISE observing cadence.}  

\begin{deluxetable}{c|cccc} \label{table:corr_matrix}
\tablecaption{Correlation Matrix of non-parametric measures of the third moment of the flux PDF.}
\tablewidth{0pt}
\tablehead{\nocolhead{} & \colhead{FF} & \colhead{Rel. Asym.} & \colhead{M} & \colhead{Skewness}} 
\startdata
FF & 1.000000 & -0.860607 & -0.721460 & -0.571853 \\
Rel. Asym. & -0.860607 & 1.000000 & 0.845581 & 0.593586 \\
M & -0.721460 & 0.845581 & 1.000000 & 0.783283 \\
Skewness & -0.571853 & 0.593586 & 0.783283 & 1.000000 \\
\enddata
\end{deluxetable}

\subsection{Identification of Eclipsing Binaries} \label{subsec:find_ebs}

\begin{figure*}
\centering
\includegraphics[scale=0.65]{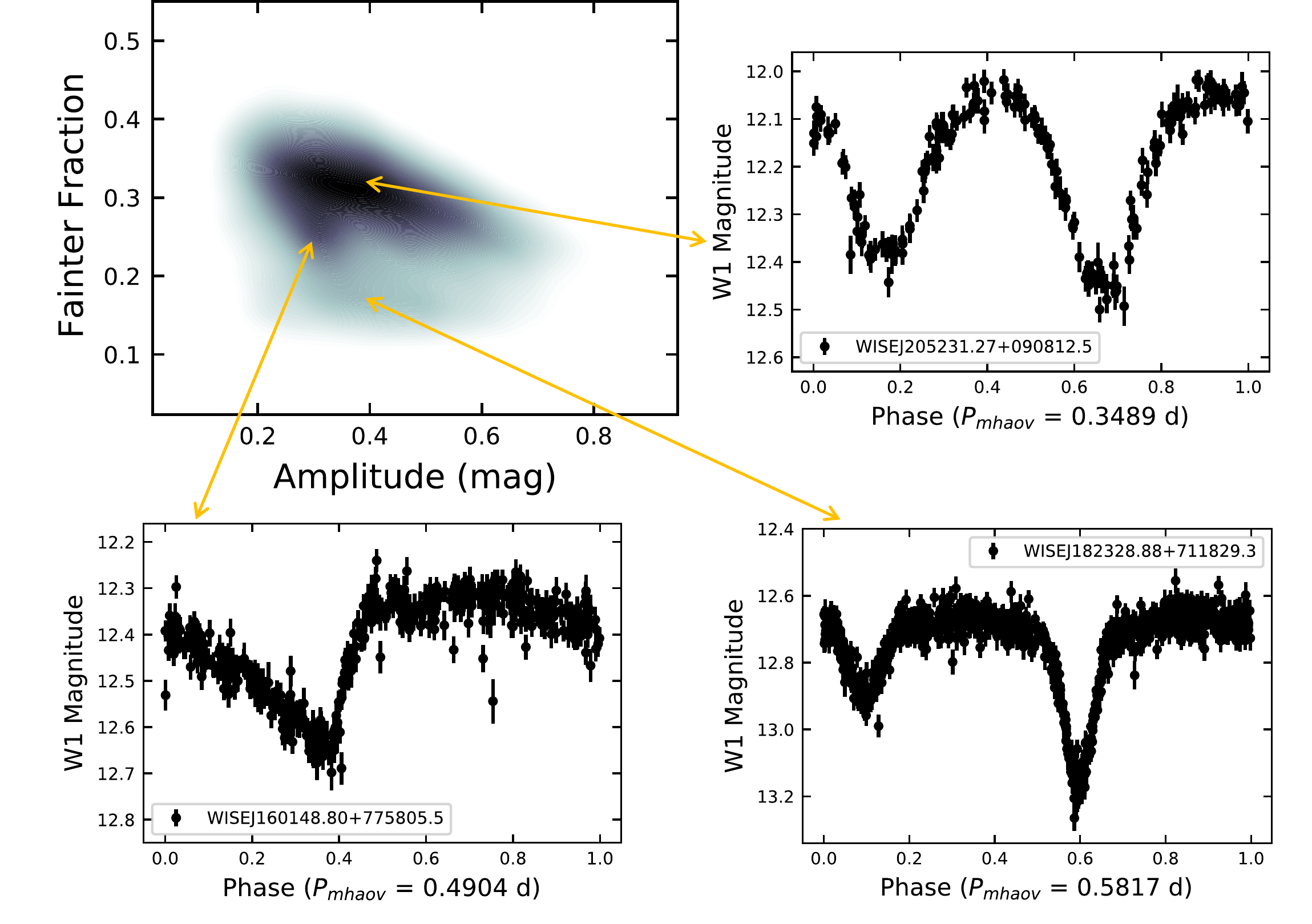}
\caption{Kernel density estimate for all periodic variables in the space of two of our non-parametric measures, \emph{fainter fraction} (\emph{FF}) and \emph{amplitude}, with example phase-folded light curves characteristic of three regions. Clockwise from top right: a contact EB found in the upper sequence phase-folded with period \edit1{$P = 2\times P_{mhaov}$}, a detached EB found in the lower horizontal sequence phase-folded with period \edit1{$P = 2\times P_{mhaov}$}, a candidate RR Lyrae found in the offshoot from the upper sequence.  The primary mechanism driving changes in luminosity for RR Lyrae is different in the optical, where it is temperature variations, as opposed to the infrared, where it is radial pulsations \citep{mado13, brag19}.  Optical and infrared light curves of RR Lyrae have different morphology and can have different skewness/\emph{FF}.}
\label{fig:eg_lc_w_ff_amp}
\end{figure*}

We identify a few physically-motivated cuts on period and a few of our non-parametric measures that can reliably isolate eclipsing binaries from other types of variability.  In Figure \ref{fig:eg_lc_w_ff_amp} we show the kernel density estimate of our periodic variable candidates in the space of \emph{fainter fraction} and \emph{amplitude}.  The sources cluster in this space and different clusters are characterized by light curves with different morphology.  Nearly all of the periodic variables have \emph{FF}$<$0.5 indicating that they are occulting.  There is an upper sequence at \emph{FF} of $\sim$ 0.35 and \emph{amplitude} between 0.2 and 0.7 that, based on visual inspection, seems to be dominated by near-contact and contact binaries.  A lower sequence at \emph{FF} of $\sim$ 0.15 and of similar amplitudes has detached binaries with more narrow eclipses. The offshoot from the upper sequence at \emph{amplitude} $\sim$ 0.3 mag and \emph{FF} $\sim$ 0.25 seems to contain RR Lyrae.

\begin{figure*}
\centering
\includegraphics[scale=1.0]{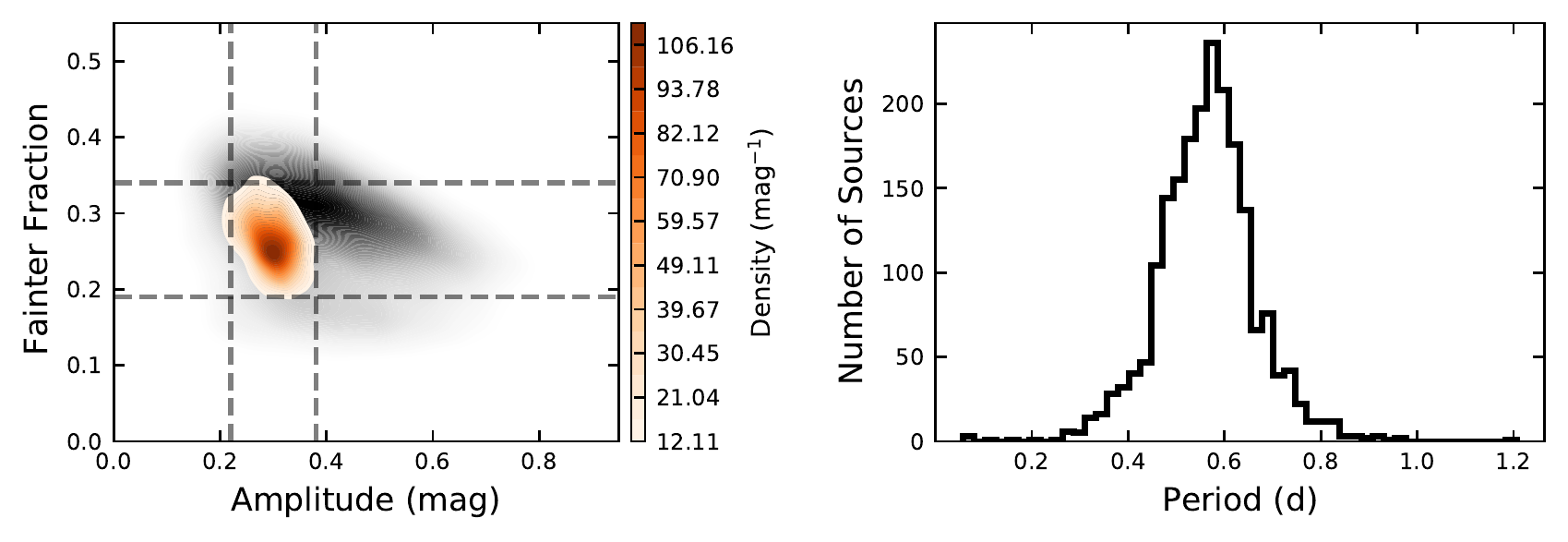}
\caption{{\bf Left: }Two-dimensional kernel density estimate for Gaia RR Lyrae in the space of two of our non-parametric measures, \emph{fainter fraction} (\emph{FF}) and \emph{amplitude}. The distribution of all the periodic variables in this space is shown in gray scale.  This plot shows that the RR Lyrae do indeed cluster in the area of the \emph{FF} vs. \emph{amplitude} plot that was suggested by Figure \ref{fig:eg_lc_w_ff_amp}. The gray, dotted lines show our initial selection of RR Lyrae.  {\bf Right: } Our period measurements for these sources.  As expected for RR Lyrae, the periods are nearly all between 0.2 days and 1 day.}
\label{fig:ff_amp_gaia_rr}
\end{figure*}

We cross-match our sample with the Gaia RR Lyrae catalog of \citet{clem19} and find 2,027 matches.  Figure \ref{fig:ff_amp_gaia_rr} shows the kernel density estimate of the cross-matched RR Lyrae in the \emph{fainter fraction}-\emph{amplitude} space and the period distribution of the RR Lyrae.  To identify candidate RR Lyrae among WISE periodic variables, we apply the cut shown by the gray, dotted lines.  Specifically, we require that 0.19 $<$ \emph{fainter fraction} $<$ 0.34 and 0.22 $<$ \emph{amplitude} $<$ 0.38.  These cuts select the entirety of the aforementioned offshoot from the upper sequence and also a portion of the upper sequence itself.  To ensure that we are not excluding too many close EBs, we restrict the \edit1{MHAOV periods} to between 0.25 and 1 days. After applying this cut, both the upper and lower sequences disappear and the RR Lyrae offshoot becomes the dominant feature in the  \emph{fainter fraction}-\emph{amplitude} space.  We choose a  lower period bound of 0.25 days because we do not expect many RR Lyrae with periods \edit1{below 0.25 days} and we do not want to exclude a high number of close eclipsing binaries.  This figure shows that these three variables - period, \emph{amplitude}, and \emph{fainter fraction} - provide a means of separating RR Lyrae from EBs without resorting to the color-magnitude diagram.

All told, this selection labels 6,100 of the periodic variables as candidate RR Lyrae.  Included in this are 1,752 out of the 2,027 ($\sim 86\%$) Gaia RR Lyrae from the cross-matched catalog of \citet{clem19}.  We exclude the remaining Gaia RR Lyrae from the EB sample as well.  In addition, we also cross-match our periodic variables with the Gaia DR2 single-object-study Cepheid catalog \citep{clem19}.  As mentioned above, we expect few Cepheids in the sample.  The cross-match reveals only 347 Gaia Cepheids and we remove these from the EB sample.

The above cuts remove 1,801 out of 2,319 ($\sim$78\%) of the RR Lyrae identified by \citet{chen18} while retaining 29,483 out of the 32,151 sources ($\sim$92\%) classified by \citet{chen18} as EBs.  We add these excluded sources back in to our EB sample.  We remove the remaining RR Lyrae and also the 710 Cepheids classified by \citet{chen18}.  We are left with a low contamination sample of 50,722 eclipsing binary candidates.

\section{Discussion}\label{sec:disc}
 
\subsection{Main-Sequence Binaries}
 
We next turn to the color-magnitude diagram to isolate main-sequence EBs.  Of our initial sample of 50,722 EB candidates, 44,550 have a Gaia best-neighbor cross-match.  Applying the quality cuts of Section \ref{sec:cont} leaves us with 32,929 sources.  To select main-sequence binaries, we start with the \citet{hame19} Pleiades spline fit shown in Figure \ref{fig:HR_ms_spine}.  Binaries are expected to be found above the main sequence for a wide range of mass ratios \citep{hurl98}.  \edit1{We saw in Section \ref{sec:cont} that our periodic variables, and especially those on the main sequence, were dominated by EBs, and the high density of sources above the main-sequence fit can be seen in Figure \ref{fig:HR_ms_spine}.}  We require that the source have an absolute G-band magnitude above the Pleiades spline fit and within 1.5 magnitudes of the spline fit value for a given color. \edit2{This cut is motivated by the fact that doubling the flux corresponds to a magnitude increase of $\sim$ 0.75.  We double this value to incorporate different stellar metallicities and measurement uncertainties.}  This cut leaves 21,746 sources. 

\begin{figure}
	\includegraphics[scale=1.0]{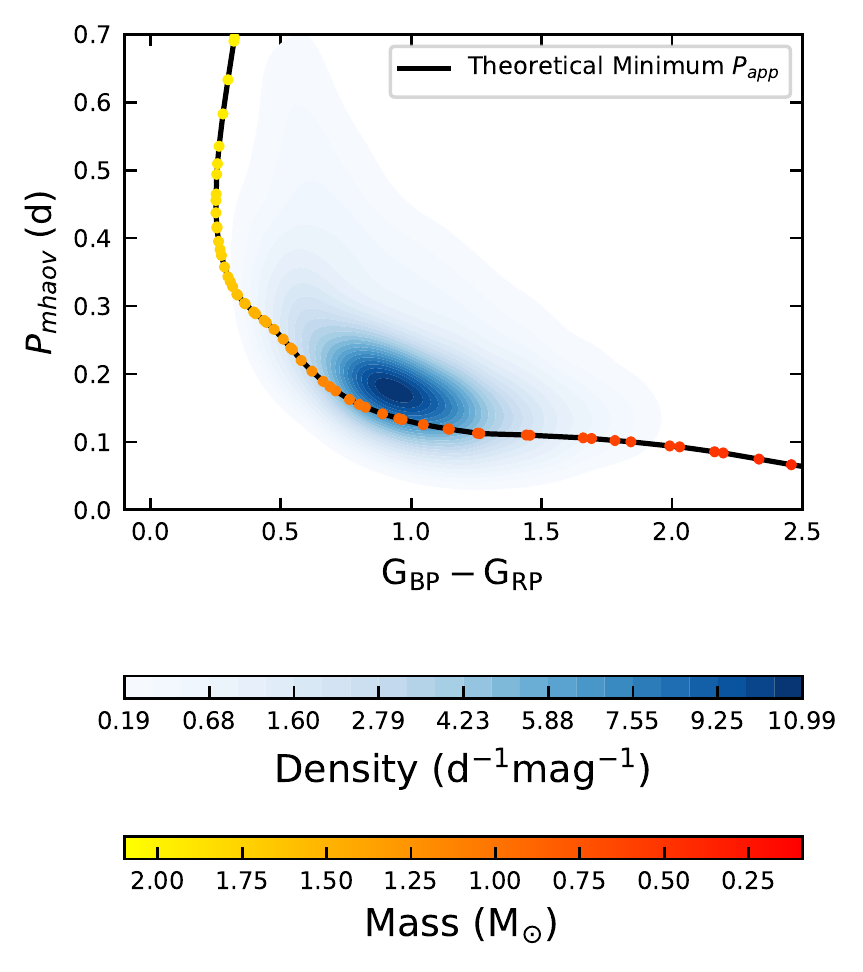}
	\caption{Gaussian kernel density estimate in the space of \edit1{MHAOV periods (which we take to be $0.5 P_{orbital}$)} and Gaia BP-RP color for MS binaries.  The black line overlaid with the red-yellow points represent the theoretical minimum apparent period ($0.5P_{orbital}$) for a contact, equal-mass, main-sequence binary.}
	\label{fig:per_color}
\end{figure}

Figure \ref{fig:per_color} shows the distribution of \edit1{MHAOV period} versus Gaia BP-RP color for our main-sequence EBs detected with WISE.  The distribution is smooth across the entire period range and includes periods not probed by \citet{chen18}.  \edit1{To make this plot, we make the approximation that $P_{mhaov} = 0.5P_{orbital}$, that is, the periodogram identifies half the orbital period.  In our cross match with the results of \citet{chen18}, this is valid for $\sim$90\% of the binaries in our sample.  However, many of our periodic variables that were not detected by \citet{chen18} have shorter periods (i.e. $P_{mhaov} < 0.15$ d) and we expect these shorter period binaries to be more likely to have two similar minima per cycle.  This could increase the percentage of sources for which we identify half the orbital period to $>$90\%.  The remaining sources for which we identify the actual orbital period can be found further from the black line in the upper part of Figure \ref{fig:per_color}.  For binaries with two identical minima per cycle, we call half the orbital period the apparent period ($P_{app} = 0.5P_{orbital}$).}  In blue is a kernel density estimate and the black line represents the theoretical minimum possible apparent period for an equal-mass, contact, main sequence eclipsing binary of a fixed age (1 Gyr) and solar metalicity as a function of mass and color.  The calculation of this line follows \citet{hwan20}.  Briefly, from the PARSEC isochrone \citep{bres12} for an age of 1 Gyr and solar metalicity we get the stellar radius of an undistorted star (i.e. the Roche lobe volume radius, $R_L$). Following \citet{eggl83}, we use the relationship between $R_L$ and the semi-major binary axis, $a$, to solve for $a$ ($R_L$ = $0.38a$).  Finally, we set the masses of the two constituent stars, $M_1$ and $M_2$, equal to each other and solve for the minimum possible apparent period using 
\[P_{app} = 0.5P_{orbital} = \pi\left(\frac{a^3}{G(M_1 + M_2)}\right)^{\frac{1}{2}}\] 
where $G$ is the gravitational constant. 

Many of our EBs are near this theoretical lower bound indicating that the majority of these systems are contact or near-contact binaries.  Interestingly, some of the points are below the black, minimum-period line.  These are systems for which the input assumptions break down.  They can be systems in which the component stars are not of equal mass, not of identical color, younger than 1 Gyr, or of lower-than-solar metallicity.  Some RR Lyrae are located close to the main sequence and may be present on the blue side of the plot ($G_{BP}-G_{RP}<0.6$) accounting for some of the spillover in the bottom left.  The dearth of sources hugging the line in the upper-left of the plot does not appear be due to decreased sensitivity to those periods (see subsection \ref{subsec:complete}). It is possible that there are physical reasons for the paucity of contact binaries in the corresponding mass range -- for example, if the magnetic braking mechanism is responsible for creating contact binaries \citep{hwan19b}, then it is expected to be inefficient at these masses due to lack of convection at $M>1.3M_{\odot}$ \citep{matt11}.

The peak in the period distribution of our main-sequence EBs is located about at \edit1{$2P_{mhaov}\approx 0.34$ days}. This is higher than the maximum of the period distribution of $\sim$0.27 days found by \citet{ruci07}, but our sample is not volume-limited. \edit1{Also, as mentioned above, for $\lesssim10\%$ of these binaries, we have detected the actual orbital period so multiplying $P_{mhaov}$ by two creates a tail to higher periods.}

\subsection{Shortest-Period Objects}

\begin{figure*}
    \centering
	\includegraphics[scale=1.0]{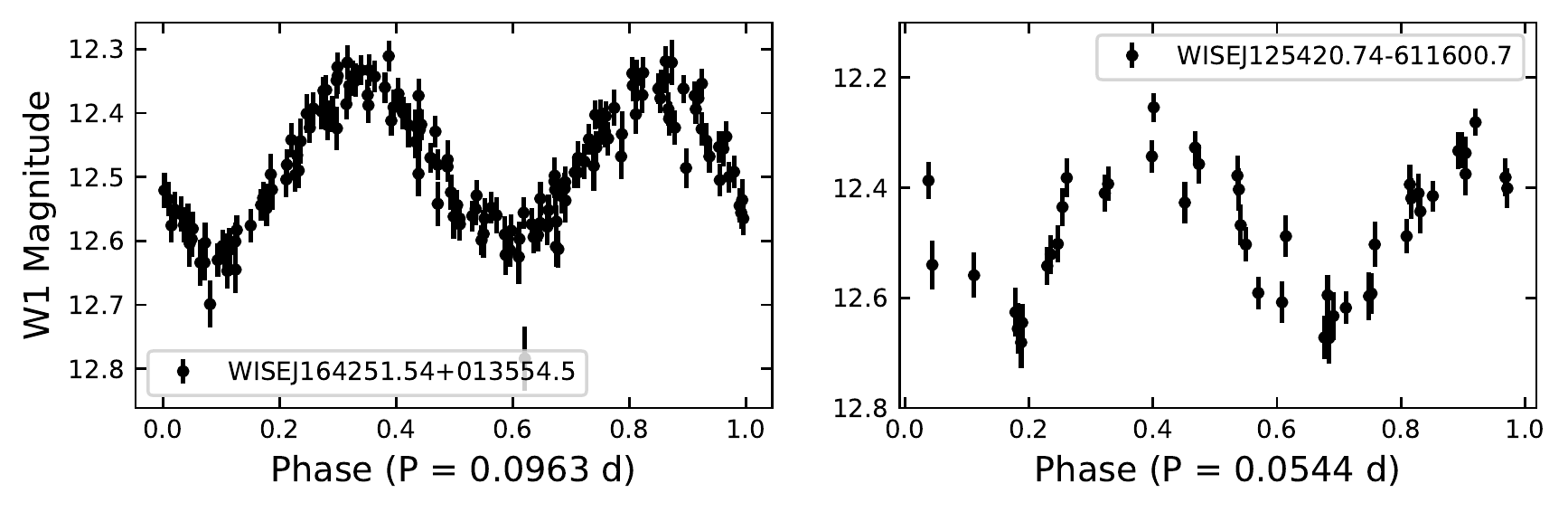}
	\caption{Phase-folded light curves for a couple of the shortest-period binaries found by our search.  The period, $P$, listed is the total period of oscillation used to fold the light curves}
	\label{fig:ultra_short}
\end{figure*}

There are 126 sources in our eclipsing binary candidates that have \edit1{MHAOV periods} less than or equal to 0.1 days. We expect higher contamination and lower period accuracy in this period range due to the limitations of the WISE cadence.  We visually inspect the phase-folded light curves and find four are the result of bad data or an erroneous period measurement and that the remaining 122 represent real periodic signals.  Figure \ref{fig:ultra_short} shows some example phase-folded light curves for these objects. These sources are prime candidates for comparison with other short-period binary catalogs (e.g \citealt{nort11, drak14b}) and subsequent analysis with the tools of \citet{conr20}.  We cross-match these sources with SIMBAD with a matching distance of 1 arcsecond. After removing the bad data, we find 37 matches of which 31 were classified in SIMBAD as a specific type of variability.  Of these 31, we find that 4 are pulsating and the remaining 27 are all classified as some sort of binary.  Three of these sources are cataclysmic variables (V* BL Hyi; SDSS J121209.30+013627.7; RX J2218.5+1925).  We find compatible periods for all of these cataclysmic variables indicating that our period measurements are accurate even in this short-period regime \citep{schm05,thor09, avva13}.  In addition, five sources are ellipsoidal variables indicating that the periodicity is due to the gravitational distortion of the stars \citep{morr85}.

\subsection{Young Stellar Objects and Extra-Galactic Variables}

We compare $R$ values and \emph{amplitudes} for blazars and young stellar objects (YSOs).  We cross-match all variable sources in our sample with the blazar catalogs of \citet{dabr19} and the YSO catalogs of \citet{mart16} as mentioned in subsection \ref{subsec:gaia}.  We find 44,196 YSOs of which 237 were classified as periodic and 981 blazars of which 2 were classified as periodic. Seven sources classified both as blazars and YSOs are removed from subsequent analysis.  

\begin{figure*}
\centering
\includegraphics[scale=1.0]{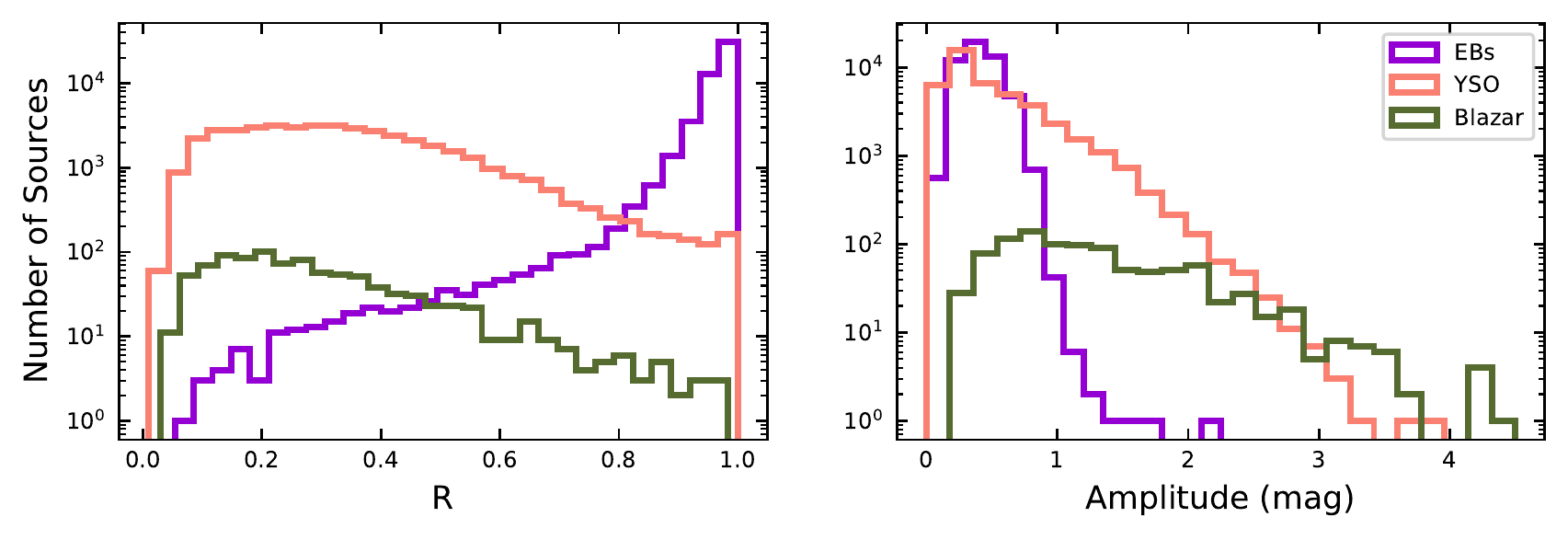}
\caption{{\bf Left:} \edit1{Ratio of variation amplitudes in the short to long terms}, $R$, for cross-matched young stellar objects and blazars. EBs are shown for comparison. {\bf Right:} \emph{Amplitude} for cross-matched young stellar objects and blazars}
\label{fig:yso_blaz}
\end{figure*}

In Figure \ref{fig:yso_blaz} we show the results of this comparison.  The distributions for $R$ are largely similar with peaks around $\sim$ 0.2-0.3 indicating that both YSOs and blazars tend to vary on longer timescales.  For \emph{amplitude}, the YSOs are peaked below 2 magnitude while the blazar distribution is more uniform. 

\begin{figure*}
\centering
\includegraphics[width=\textwidth]{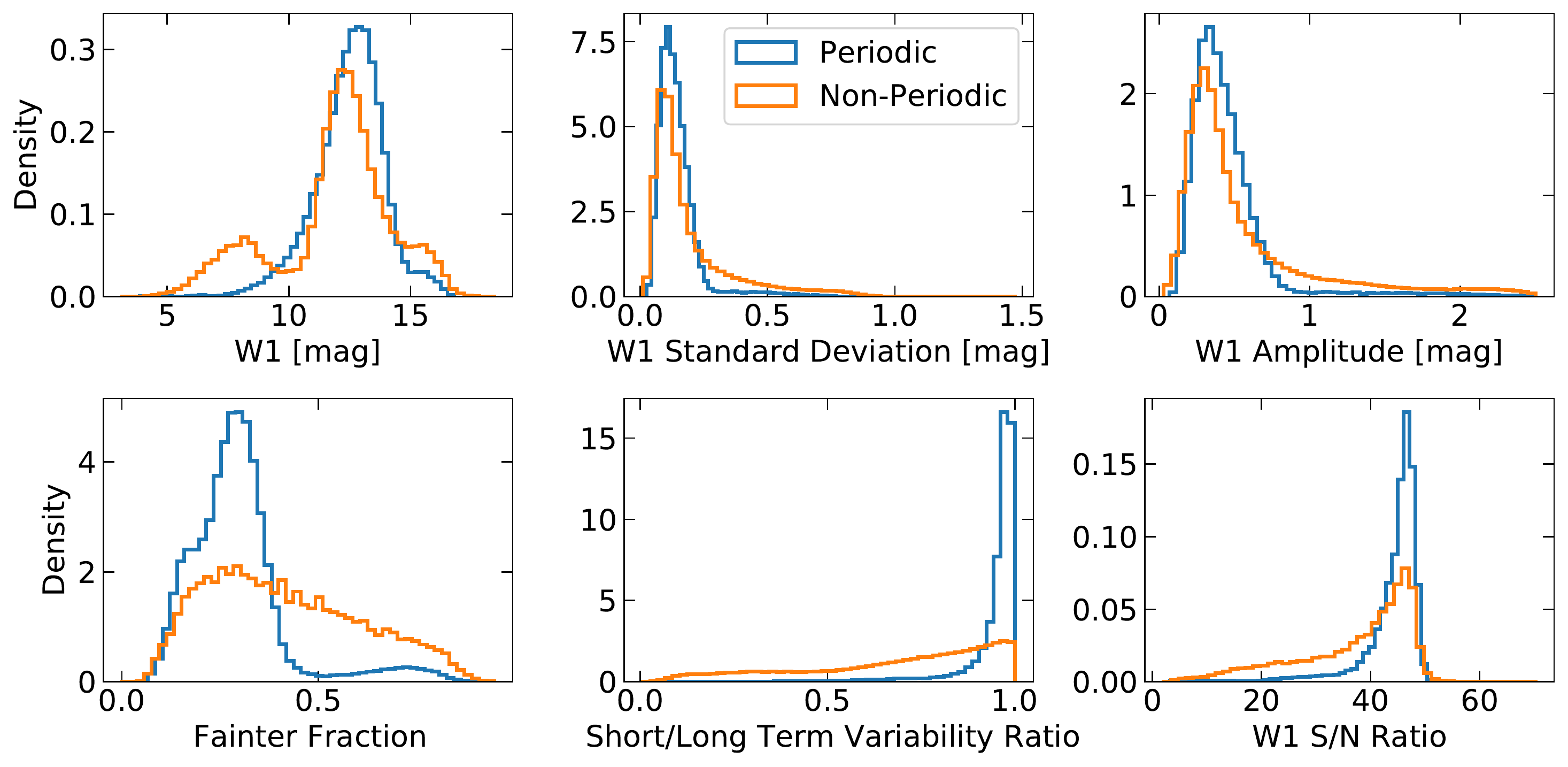}
\caption{Distribution of periodic and aperiodic variables for a variety of non-parametric measures. The peak at 8th magnitude in the W1 magnitude plot is caused by saturation, rather than by physical variability, which is why there is not a corresponding peak in the distribution of periodic variables \citep{niku14}.  We remove these sources as well as very faint sources with the quality cut of Section \ref{subsec:selection}.  The \emph{fainter fraction} plot shows that our sample is dominated by periodic variables of the occulting type. The histograms are normalized to have unit area, to account for the fact that our sample contains more aperiodic variables than periodic variables.}
\label{fig:non_parametric_plts}
\end{figure*}

\subsection{Separation of Periodic and Aperiodic Variables with Non-Parametric Features}

Thus far, we have used our non-parametric measures for the classification of periodic variables identified with a periodogram.  In this section, we discuss further applications of these measures.  In particular, we show that the non-parametric features contain the requisite information to identify periodic sources.

Figure \ref{fig:non_parametric_plts} shows the distribution of periodic and aperiodic sources for a variety of the non-parametric measures. We additionally show the distribution of W1 signal-to-noise (S/N) ratio, since the quality of the data can influence the determination of periodicity. For many of these measures (especially the ratio of short-term to long-term variability), there is a marked difference between the distributions of periodic and aperiodic variables. 

We investigate the potential of discriminating periodic variables from aperiodic variables using non-parametric features alone. Using empirical testing, we select the five most informative features -- the W1 magnitude, W1 signal-to-noise ratio, estimated variability \emph{amplitude}, \emph{fainter fraction}, and the ratio between long-term and short-term variability. The variability \emph{amplitude} and standard deviation have similar distributions (Figure \ref{fig:non_parametric_plts}), but we find that the \emph{amplitude} is a marginally better discriminator.  We apply the quality cuts of Section \ref{subsec:selection} on the W1 magnitude to exclude saturated and very faint sources and to ensure that we are not classifying trivially based on this data quality metric.

We define a two-class classification problem where the `positive' class is defined as periodic variability, and the `negative' class corresponds to aperiodic variability. We consider two classification models -- the logistic regression and the random forest. In both cases, the general framework is the same -- the model coefficients are solved for by `training' on stars with known classes. The models can then perform predictions on new data, returning a probability $\hat{p}$ that a given star exhibits periodic variability. The logistic regression model assumes a linear relation between the features and the log-odds of a star exhibiting periodic variability \citep{yu11}. The random forest assumes no parametric model, instead relying on logical decision trees to map the feature space to the true and false classes \citep{brei01}.

\edit1{To evaluate our methods, we randomly select 75\% of the stars in our sample for training, and make predictions on the remaining validation stars (25\%) that were left out of training. In both the train and test samples, we randomly downsample the number of aperiodic stars to make it balanced with the number of periodic stars. Repeating our experiment for different random selections (or different ratios of train/test sample size) does not significantly affect our results. We find that, on average, the classifier correctly classifies stars as aperiodic or periodic 90\% of the time with the logistic regression, and 94\% with the random forest.} The random forest model outperforms the logistic regression, likely because a linear model is insufficient to describe the relationships between our features. A key metric for our particular use-case is precision -- the probability that a star classified as periodic is genuinely periodic. A higher precision means a lower false positive rate, preventing wasted follow-up resources. We find that the random forest model trained on the five non-parametric features achieves an average precision of 97\% on our test data, an improvement over the logistic regression's 91\%. \edit1{The confusion matrix of the predictions on the test dataset is as follows, with true aperiodic and true periodic as the columns from left to right, and classified aperiodic and classified periodic as rows from top to bottom:}

\begin{equation*}
\begin{bmatrix}
    11859 & 617 \\
    856 & 12098
\end{bmatrix}
\end{equation*} \edit1{As is desirable, most objects fall on the diagonal of the confusion matrix, and the number of false positives and false negatives is similar.}

An interesting application of these tools is to use the probabilities returned by the classifier to rank interesting candidates. The probabilities themselves can be incorporated into a hierarchical search that uses other prior information to inform the selection. In future large-scale surveys, such a pre-selection will be essential to efficiently allocate computational resources by running the full periodogram analysis on high-confidence periodic candidates first. 

The brief demonstration of this section is intended mainly as a proof-of-concept of the information content of these features and has some important caveats.  The first caveat is that one of the features (S/N ratio) measures data quality while others deal with real information about the source from the flux PDF and timescales of variability.  The discrimination based on data quality is trivial; without good data it is \edit1{difficult} to flag a source as periodic. Therefore, it is important to ensure that the classifier is not biased to predict periodicity when the data quality is high. \edit1{First, we quantify the relative importance of the input features. The random forest classifier has a built-in metric that ranks the importance of the input features, based on how much that feature contributes to the prediction quality \citep{brei01}. The features ordered from most to least important are the short-to-long term variability ratio, W1 signal-to-noise ratio, fainter fraction, estimated amplitude, and W1 magnitude.} 

Including the S/N ratio as a feature improves the classification accuracy by $\sim\ 5\%$ compared to if solely the W1 magnitude is used. We interpret this as the S/N ratio breaking a degeneracy in the variability measures between noisy sources and truly variable sources -- objects with low S/N will automatically display variability due to noise, so including the S/N ratio as an explicit feature enables the classifier to incorporate this information. As another check, we run the classification on only high S/N sources, and yield nearly identical performance. Therefore, the classification appears to rely on meaningful information about the source, and does not seem to be biased by data quality.

A second caveat is that this demonstration is limited by the nature of the training sample.  For example, due to the WISE cadence, long-period variables are almost completely absent from our training sample and so we do not expect the classifier to be adept at identifying them.  Also, the use of periodogram-based classification as the `ground-truth' for training biases the classifier to be more likely to detect similar types of periodicity as the periodogram, albeit at a much lower computational cost.

That said, the non-parametric approach has some important advantages over the periodogram and can even yield valuable information without reference to a periodogram-based analysis \citep[e.g.,][]{rimo19}.  PDF measures like this are more robust to photometric errors than the periodogram, and are meaningful even in systems with changing periods.  These systems are of great interest, but would not normally be detected by the periodogram because implicit in the periodogram approach is the assumption of a constant period.

\section{Conclusions}  \label{sec:conc}

In this paper we present an analysis of $\sim$450,000 WISE variables. The variables are identified based on their r.m.s. variability in the AllWISE survey \citep{hoff12}. \edit1{We combine their AllWISE and NEOWISE light curves and apply quality cuts to filter out low-quality individual exposures.}

\edit1{Using these combined light curves, we conduct periodogram analysis to identify $\sim$56,000 periodic variables, $\sim$19,000 of which were not included in the previous WISE periodic variable catalog of \citet{chen18}.  We search for periodicity over a finely spaced grid in frequency between $0.1$ and $20$ days$^{-1}$ and use the multi-harmonic analysis of variance (MHAOV) periodogram with 5 model parameters.  This periodogram models the light curve as a series of periodic, orthogonal functions and uses a variance statistic to quantify the quality of the fit at each test frequency.  To classify an object as periodic, we require a significant value of the statistic at a frequency non-coincident with that of the WISE observing cadence, a sufficiently small period uncertainty to prevent a large phase offset over the time-series duration, and  good phase coverage.  We cross-match with Gaia DR2 and use the distribution of our periodic variables in the space of color-absolute magnitude to get a sense of our catalog contents.  We find that the sample is dominated by eclipsing binaries.}

\edit1{Next, we compute a variety of non-parametric variability measures for all of the $\sim$450,000 WISE variables and show that these measures are useful both for (a) separating periodic and aperiodic variables and (b) classifying the type of periodicity.  These non-parametric measures include a variety of measures of the flux distribution range and moments.  We introduce a new measure of the third moment and compare it to previously existing measures.  These non-parametric measures also include an estimate of the timescale of variability ($R$, the ratio of variability amplitude on day timescales to that on month/year timescales).}

In terms of classification (b), we show that these interpretable and easily implemented measures provide an effective means of isolating eclipsing binaries from other types of periodic variability. We identify an all-sky sample of $\sim$51,000 eclipsing binaries in the infrared.  The majority of these binaries are contact or near-contact making them prime targets for future study.

In terms of the identification of periodic variables (a), we demonstrate the high information content of the non-parametric features by using them to identify periodic variables at a much lower computational cost than the traditional, periodogram-based analysis.  This type of analysis can be used to speed future studies of periodic variability and, in some cases, bypass the periodogram-based analysis entirely (e.g. \citealt{rimo19}).  Furthermore, the non-parametric method overcomes some of the short-comings of periodograms.  Importantly, because it does not implicitly assume a constant signal period as the periodogram does, it offers the possibility of identifying binaries exhibiting orbital evolution on human timescales.

\textbf{Data Availability:} The full catalog of WISE variables, periodic variables, and binaries is available as an electronic supplement to this paper\footnote{\url{https://zakamska.johnshopkins.edu/data.htm}}.  The data model is listed in Table \ref{table:data_model}\edit1{.}

\acknowledgments
{The authors are grateful to the anonymous referee for the constructive feedback.  This publication makes use of data products from the Wide-field Infrared Survey Explorer, which is a joint project of the University of California, Los Angeles, and the Jet Propulsion Laboratory/California Institute of Technology, funded by the National Aeronautics and Space Administration. This work has made use of data from the European Space Agency (ESA) mission {\it Gaia} (\url{https://www.cosmos.esa.int/gaia}), processed by the {\it Gaia} Data Processing and Analysis Consortium (DPAC, \url{https://www.cosmos.esa.int/web/gaia/dpac/consortium}). Funding for the DPAC has been provided by national institutions, in particular the institutions participating in the {\it Gaia} Multilateral Agreement. This research made use of the cross-match service provided by CDS, Strasbourg. This research has made use of the VizieR catalogue access tool, CDS, Strasbourg, France.  EP was supported in part by the Provost's Undergraduate Research Award at Johns Hopkins University. EP and HCH acknowledge support by the Space@Hopkins seed grant. The authors acknowledge support from NASA ADAP grant 80NSSC19K0581.  The MHAOV periodogram and the orthogonal, trigonometric function fit are implemented with the time series analysis package of Alex Schwarzenberg-Czerny that is based on a wrapper scheme by Ewald Zietsman.  This research made use of Astropy, a community-developed core Python package for Astronomy \citep{astr13, astr18}.}

\bibliography{AstroBase}
\bibliographystyle{aasjournal}

\end{document}